\documentclass[11pt, a4paper]{article}
\pdfoutput=1
\usepackage{graphicx}
\usepackage{amssymb}
\usepackage{amsmath}
\usepackage{bm}
\usepackage{url}
\usepackage[table]{xcolor} %for \rowcolors
\usepackage{cite}
\usepackage{slashed}
\usepackage{subfigure}
\usepackage{epstopdf}            
\usepackage{epsfig}
\usepackage{here}
\usepackage{comment}
\usepackage{xfrac} % Nice small fractions \sfrac
\usepackage{multirow}
\usepackage{adjustbox}
\usepackage{booktabs} %for \toprule
\usepackage{colortbl} %for \rowcolor
\usepackage{lmodern} %for font warning

\renewcommand{\thefootnote}{\fnsymbol{footnote}}
\numberwithin{equation}{section} % Eq.(Sec.eq.)
\def\beq#1\eeq{\begin{align}#1\end{align}}
\newcommand{\ov}{\overline}

\renewcommand{\arraystretch}{1.3}

\newcommand{\eg}{{\em e.g.}}
\newcommand{\ie}{{\em i.e.}}

\RequirePackage{xspace}
\def\Bbar    {\kern 0.18em\overline{\kern -0.18em B}{}\xspace}
\def\Bb      {\ensuremath{\Bbar}\xspace}

\def\Jpsi   {J\mbox{\footnotesize$\!/\psi$}}

\usepackage{caption}
\definecolor{BlueViolet}{rgb}{0.2, 0.00, 0.7}
\definecolor{Blue}{rgb}{0.15, 0.00, 0.9}
\definecolor{light_blue}{rgb}{0.15, 0.35, 0.95}
\definecolor{kit_green}{rgb}{0, 
0.58823 %150/255
, 0.50980 %130/255
}
\usepackage[%dvipdfmx,
colorlinks=true, linkcolor=light_blue,citecolor=light_blue,urlcolor=kit_green]{hyperref} 
\usepackage[hyphenbreaks]{breakurl} % for breaking long URLs on arxiv.org

\normalsize %space between the lines %same as JHEP

\begin{document}
\sloppy %for good hyphenation
%https://tex.stackexchange.com/questions/9107/how-can-i-make-my-text-never-go-over-the-right-margin-by-always-hyphenating-or-b

\begin{titlepage}

\begin{center}
\begin{flushright}
KEK--TH--2614, P3H--24--024, TTP24--008, CHIBA--EP--262\\ 
\end{flushright}

\vskip .4in

{\fontsize{16pt}{0cm}\selectfont
 {\bf  Global fit to 
$\boldsymbol{b \to c\tau\nu}$ anomalies as of Spring 2024
}
}

\vskip .2in

{\large 
{\bf Syuhei Iguro,$^{a,b,c,d,e}$
Teppei Kitahara,$^{b,f,g}$
and 
Ryoutaro Watanabe$^{h}$}}

\vskip .3in

\begingroup\small
\begin{tabbing}
$^{a}$ \= {\it 
Institute for Advanced Research, Nagoya University, Nagoya 464--8601, Japan}
\\[0.2em]
$^{b}$ \> {\it 
Kobayashi-Maskawa Institute for the Origin of Particles and the Universe, Nagoya University,} \\
\> {\it Nagoya 464--8602, Japan}
\\[0.2em]
$^{c}$ \> {\it 
KEK Theory Center, IPNS, KEK, Tsukuba 305--0801, Japan} 
\\[0.2em]
$^{d}$ \> {\it 
Institute for Theoretical Particle Physics (TTP), Karlsruhe Institute of Technology (KIT),} \\
\> {\it Wolfgang-Gaede-Str.\,1, 76131 Karlsruhe, Germany}
\\[0.2em]
$^{e}$ \> {\it 
Institute for Astroparticle Physics (IAP),
Karlsruhe Institute of Technology (KIT), } \\
\> {\it Hermann-von-Helmholtz-Platz 1, 76344 Eggenstein-Leopoldshafen, Germany}
\\[0.2em]
$^{f}$ \> {\it 
Department of Physics, Graduate School of Science,
Chiba University, Chiba 263--8522, Japan} 
\\[0.2em]
$^{g}$ \> {\it 
CAS Key Laboratory of Theoretical Physics, Institute of Theoretical Physics,} \\
\> {\it Chinese Academy of Sciences, Beijing 100190, China}
\\[0.2em]
$^{h}$ \> {\it 
Institute of Particle Physics and Key Laboratory of Quark and Lepton Physics (MOE),} \\
\> {\it Central China Normal University, Wuhan, Hubei 430079, China}
\end{tabbing}

\vskip 0.05in
{\href{mailto:igurosyuhei@gmail.com}{igurosyuhei@gmail.com}, \href{mailto:kitahara@chiba-u.jp}{kitahara@chiba-u.jp}, \href{mailto:watanabe@ccnu.edu.cn}{watanabe@ccnu.edu.cn (corresponding author)}}\\
\endgroup 
\vskip 0.075in
{Dated: October~7, 2024} 
\end{center}
%\vskip 0.05in

\begin{abstract}
\noindent 
%%%
Recently, several new experimental results of the test of lepton flavor universality (LFU) in $B\to D^{(\ast)}$ semi-leptonic decays were announced:
the first result of $R_{D}$ from the LHCb Run\,1 data,
the first results of $R_{D}$ and $R_{D^\ast}$ from the LHCb Run\,2 data, and the first result of $R_{D^\ast}$ from the Belle~II collaboration.
Including these new data, a global analysis still prefers the violation of the LFU between the tau and light leptons.
A new world average of the data from the BaBar, LHCb, Belle, and Belle~II collaborations is 
$R_{D} = 0.342 \pm 0.026$ and
$R_{D^{\ast}} = 0.287 \pm 0.012$.
Including this new data, we update a circumstance of the $b \to c \tau \overline\nu$ measurements and their implications for new physics (NP).
Incorporating recent developments for the $B \to D^{(\ast)}$ form factors in the Standard Model (SM),
we observe a $4.4 \sigma$ deviation from the SM prediction. 
Our updates also include; 
model-independent NP formulae for the related observables; 
and the global fittings of parameters for leptoquark scenarios as well as single NP operator scenarios. 
Furthermore, we show future potential to indirectly distinguish different NP scenarios 
with the use of the precise measurements of the polarization observables in $B\to D^{(\ast)}\tau \overline\nu$ at the Belle~II and 
the high-$p_{\rm T}$ flavored-tail searches at the LHC. 
We also discuss an impact on the LFU violation in $\Upsilon \to l^+ l^-$.

%%%%%%%%%%%%%%%%%%%%%%%%%
\end{abstract}
{\sc Keywords:} 
Beyond Standard Model, $B$ physics, 
Effective Field Theories
%%%%%%%%%%%%%%%%%%%%%%%%%
\end{titlepage}

\setcounter{page}{1}
\renewcommand{\thefootnote}{\#\arabic{footnote}}
\setcounter{footnote}{0}

%%%%%%%%%%%%%%%%%%%%%%%%%
% Contents
%%%%%%%%%%%%%%%%%%%%%%%%%
\hrule
\tableofcontents
\vskip .2in
\hrule
\vskip .4in

%%%%%%%%%%%%%%%%%%%%%%%%%

%%%%%%%%%%%%%%%%%%%%%%%%%%%%%%%
\section{Introduction}
\label{sec:intro}
%%%%%%%%%%%%%%%%%%%%%%%%%%%%%%%

The semi-tauonic $B$-meson decays, $\Bb \to D^{(\ast)} \tau \overline{\nu}$, have been intriguing processes to measure the lepton flavor universality (LFU):
 \begin{align}
  R_D \equiv \frac{\mathcal{B}(\Bb\rightarrow D \,\tau\, \overline\nu_\tau)}{\mathcal{B}(\Bb\rightarrow D\, \ell\,\overline\nu_\ell)} \,, \qquad
  R_{D^{\ast}} \equiv \frac{\mathcal{B}(\Bb\rightarrow D^{\ast} \tau \,\overline\nu_\tau)}{\mathcal{B}(\Bb\rightarrow D^{\ast} \ell\,\overline\nu_\ell)} \,,
\end{align}
since it has been reported that the global average of the measurements by 
the BaBar~\cite{Lees:2012xj,Lees:2013uzd},
LHCb from Run\,1 and Run\,2 data~\cite{Aaij:2015yra,Aaij:2017uff,Aaij:2017deq,LHCb:2023zxo},
Belle~\cite{Huschle:2015rga,Hirose:2016wfn,Hirose:2017dxl,Abdesselam:2019dgh,Belle:2019rba}, 
and recently Belle~II
collaborations \cite{Belle-II:2020sdf,Belle-II:2024ami}
indicates a deviation from the Standard Model (SM) prediction. Here, $\ell =e,~\mu$ for the BaBar/Belle/Belle~II, while $\ell =\mu$ for the LHCb collaborations. 
See Table~\ref{tab:RD_exps} for the present experimental summary 
with including the HFLAV collaboration's preliminary average in the spring of 2024 \cite{HFLAV2024winter}.

A key feature of this deviation is that 
the world average of the measured values of $R_{D}$ and $R_{D^{\ast}}$ significantly exceeds their SM predictions
and thus implies violation of the LFU between the tau and light leptons. 
Then it has been followed by a ton of theoretical studies to understand its implication from various points of view, \eg, see Refs.~\cite{London:2021lfn,Capdevila:2023yhq} and references therein. 
A confirmation of the LFU violation will provide evidence of new physics (NP).
%

%%%%%%%%%%%%%%%%%%%%%%%%%%%%%%%
\subsection{Summary of the current status: Spring 2024}
%%%%%%%%%%%%%%%%%%%%%%%%%%%%%%%

The LHCb collaboration showed their results of $R_{D^\ast}$ with a semileptonic-tagging method ($\tau\to\mu\overline\nu\nu$) in 2015 \cite{Aaij:2015yra} and a hadronic-tagging method ($\tau\to 3 \pi \nu$)  in 2017 \cite{Aaij:2017uff,Aaij:2017deq} using the LHCb Run\,1 dataset.
At the end of 2022 \cite{LHCbRun1}, they reported
the first results of $R_{D}$ as well as $R_{D^{\ast}}$ with the semileptonic tagging method 
using the LHCb Run\,1 dataset \cite{LHCb:2023zxo},
\begin{align}
    \begin{aligned}
    R_D^{\rm LHCb2022} & =0.441\pm 0.060_{\rm stat} \pm 0.066_{\rm syst}\,,\\
    R_{D^\ast}^{\rm LHCb2022} &= 0.281 \pm 0.018_{\rm stat} \pm 0.024_{\rm syst}\,.
    \end{aligned}
\end{align}
This result superseded the previous result reported in 2015.
Furthermore, 
in early 2023 \cite{LHCbRun2}, 
they reported the first result of $R_{D^{\ast}}$ with the hadronic-tagging method 
using the combined dataset of the LHCb Run\,1 and  part of the Run\,2
\cite{LHCb:2023uiv}, 
\begin{align}
    R_{D^\ast}^{\rm LHCb2023} &= 0.267 \pm 0.012_{\rm stat} \pm 0.019_{\rm syst}\,,
\end{align}
which superseded the previous result in 2017.

In addition, in March 2024 the LHCb collaboration announced 
preliminary results of $R_D$ and $R_{D^\ast}$ with
the semileptonic-tagging method using the partial Run\,2 dataset \cite{LHCb2024},
\begin{align}
    \begin{aligned}
 R_D^{\rm LHCb2024} & =0.249\pm 0.043_{\rm stat} \pm 0.047_{\rm syst}\,,\\
    R_{D^\ast}^{\rm LHCb2024} &= 0.402  \pm 0.081_{\rm stat} \pm 0.085_{\rm syst}\,.
    \end{aligned}
\end{align}
The uncertainty of $R_{D^\ast}^{\rm LHCb2024}$ is large, while 
the one of $R_D^{\rm LHCb2024}$ is small enough to push the world average down.

In the meantime, at last, 
the Belle~II collaboration has started their data taking from 2019 \cite{Belle-II:2020sdf}.
Recently, they reported a first preliminary result of $R_{D^\ast}$ 
with a semileptonic tagging-method  ($\tau\to\ell \overline\nu\nu$) \cite{Belle2,Belle-II:2024ami}, 
\begin{align}
R_{D^\ast}^{\rm BelleII} & = 0.262 ^{+0.041}_{-0.039}{}_{\rm stat\,} {}^{+0.035}_{-0.032}{}_{\rm syst}\,. 
\label{eq:BelleII}
\end{align}
The experimental uncertainty of the Belle~II result is still large.
However, the amount of the integrated luminosity used in the Belle~II analysis (189\,fb$^{-1}$) is only a quarter of one of the Belle (711\,fb$^{-1}$).
Therefore, it is expected that this uncertainty will be reduced significantly in the near future
 \cite{Belle-II:2022cgf}.

Moreover, the CMS collaboration has developed an innovative data recording method, called ``$B$ Parking'' since 2019 \cite{Bparking,CMS-DP-2019-043,Bainbridge:2020pgi,Takahashi}. 
Although their official first results for $R_{D^{(\ast)}}$ are still being awaited, 
it would be expected that the size of the experimental uncertainty is comparable to the other $B$ factories.

%%%Table[begin]
%%%%%%%%%%%%%%%%%%%%%%%%%%%%%%%
\begin{table}[t]
\centering
\newcommand{\bhline}[1]{\noalign{\hrule height #1}}
\renewcommand{\arraystretch}{1.5}
\rowcolors{2}{gray!15}{white}
\addtolength{\tabcolsep}{5pt} % add space between columns
   \scalebox{1}{
  \begin{tabular}{lccc} 
  %\toprule
  \bhline{1 pt}
  \rowcolor{white}
 Experiment  &$R_{D^\ast}$ & $R_{D}$ & Correlation  \\  \hline 
BaBar  
\cite{Lees:2012xj,Lees:2013uzd}
& $0.332\pm 0.024\pm 0.018$ & $0.440\pm 0.058 \pm 0.042$ & $-0.27$ \\
Belle 
\cite{Huschle:2015rga}
& $0.293\pm 0.038\pm 0.015$ & $0.375\pm 0.064\pm 0.026$ & $-0.49$\\
Belle 
\cite{Hirose:2016wfn,Hirose:2017dxl}
& $0.270\pm 0.035^{+0.028}_{-0.025}$ & --& -- \\
Belle 
\cite{Abdesselam:2019dgh,Belle:2019rba}
& $0.283\pm 0.018\pm 0.014$ & $0.307\pm 0.037\pm 0.016$ & $-0.51$\\
LHCb 
\cite{Aaij:2015yra,LHCbRun1,LHCb:2023zxo} 
& $0.281 \pm 0.018 \pm 0.024$ &$0.441\pm 0.060 \pm 0.066$ & $-0.43$ \\ 
LHCb
\cite{Aaij:2017uff,Aaij:2017deq,LHCbRun2,LHCb:2023uiv}
& $0.267\pm 0.012\pm 0.019$ & -- & -- \\
Belle\,II
\cite{Belle2,Belle-II:2024ami}
& 
$0.262 {}^{+0.041}_{-0.039} {}^{+0.035}_{-0.032}$ & --& -- \\
LHCb
\cite{LHCb2024}
& 
$0.402 \pm 0.081 \pm 0.085$ &$0.249 \pm 0.043 \pm 0.047$& $-0.39$ \\
\hline
 World average 
\cite{HFLAV2024winter} 
 & $0.287 \pm 0.012$
 &
$0.342 \pm 0.026$
&
$-0.39  % -0.3911
$ \\
\bhline{1 pt}
%\bottomrule
   \end{tabular}
}   
\addtolength{\tabcolsep}{-5pt} % set back to normal
 \caption{ \label{tab:RD_exps}
 Current status of the independent experimental $R_{D^{\ast}}$ and $R_{D}$ measurements. 
 The first and second errors are statistical and systematic, respectively. 
 }
\end{table}
%%%%%%%%%%%%%%%%%%%%%%%%%%%%%%%
%%%Table[end]

In Table~\ref{tab:RD_exps}, we summarize the current status of the $R_{D^{(\ast)}}$ measurements including the new LHCb and Belle~II results. 
It is found that the above four new results are  consistent with the previous world average evaluated 
by the HFLAV collaboration in 2021~\cite{HFLAV:2022pwe},
whose $p$-value is $28\%$ with $\chi^2/\text{dof} = 8.8/7$.
Including all available data in Table~\ref{tab:RD_exps}, the preliminary world averages of  $R_{D^{(\ast)}}$ were evaluated by the HFLAV \cite{HFLAV2024winter} as
\begin{align}
    \begin{aligned}
    R_D &= 0.342 \pm 0.026\,, \\
    R_{D^\ast} & = 0.287 \pm 0.012\,,
    \label{eq:WAvalues}
    \end{aligned}
\end{align}
with the $R_D$--$R_{D^\ast}$ correlation of $-0.39$.
It gives $p$-value among all data as $35\%$ with $\chi^2/\text{dof} = 12.11/11$.

Regarding the combined average, an important analysis is given in Ref.~\cite{Bernlochner:2021vlv}. 
The authors pointed out that evaluations of the $D^{\ast\ast}$ distributions in the SM background involve nontrivial correlations that affect the $R_{D^{(\ast)}}$ measurements. 
Their sophisticated study shows that the combined $R_{D^{(\ast)}}$ average is slightly sifted, which is beyond the scope of our work since the size of the shift depends on the details of the experimental setup.

%%%Table[begin]
%%%%%%%%%%%%%%%%%%%%%%%%%%%%%%%
\begin{table}[t]
\centering
\newcommand{\bhline}[1]{\noalign{\hrule height #1}}
\renewcommand{\arraystretch}{1.5}
\rowcolors{2}{gray!15}{white}
%\addtolength{\tabcolsep}{5pt} % add space between columns
  % \scalebox{0.8}{
   \begin{adjustbox}{width=\columnwidth,center}
  \begin{tabular}{lcccccccc} 
  %\toprule
  \bhline{1 pt}
  \rowcolor{white}
   Reference & $R_{D}$ & $R_{D^{\ast}}$ & $P_\tau^D$ & $-P_\tau^{D^\ast}$ &
   $F_L^{D^\ast}$ & $R_{\Jpsi}$ & $R_{\Lambda_c}$ & $R_{\Upsilon(3S)}$  \\  \hline 
Bernlochner, {\it et al.} \cite{Bernlochner:2022ywh} & $0.288(4)$ & $0.249(3)$ & -- & -- & -- & -- & -- & -- \\
Iguro, Watanabe \cite{Iguro:2020cpg} & $0.290(3)$ & $0.248(1)$ & $0.331(4)$& $0.497(7)$& $0.464(3)$& -- & -- & -- \\
Bordone, {\it et al.} \cite{Bordone:2019vic,Bordone:2019guc} & $ 0.298(3)$ & $0.250(3)$ & $ 0.321(3)$& $0.492(13)$& $0.467(9)$& --& --& --\\ %modified by following Bordone:2019guc
HFLAV2024 \cite{HFLAV2024winter} & $0.298 (4)$ & $0.254 (5)$ & --& --& --& --& --& --\\
Refs.~\cite{Harrison:2020nrv,Bernlochner:2018kxh,Aloni:2017eny} & -- & -- & --& --& --& $0.258(4)$& $0.324 (4)$& $0.9948$\\
\hline
Data   & $0.342 (26)$ & $0.287 (12)$ & -- & $0.38 \,^{+0.53}_{-0.55}$ & 
$0.49(5)$ &
$0.61(18)$
& $0.271 (72)$ & $0.968 (16)$ \\ 
\bhline{1 pt}
%\bottomrule
   \end{tabular}
%}
\end{adjustbox}
%\addtolength{\tabcolsep}{-5pt} % set back to normal
    \caption{\label{tab:data_SM_summary}
Summary of the SM predictions for the $\Bb \to D^{(\ast)}\tau\overline\nu$ and related observables. 
The current averaged values of the experimental measurements are also written in the last row. 
See the main text for explanations of differences between the SM predictions.
}
\end{table}
%%%%%%%%%%%%%%%%%%%%%%%%%%%%%%%
%%%Table[end]
Recent SM predictions for $R_{D^{(\ast)}}^\text{SM}$ have been obtained in Refs.~\cite{Bordone:2019vic,Bordone:2019guc,Iguro:2020cpg,Bernlochner:2022ywh,HFLAV2024winter} as summarized in Table~\ref{tab:data_SM_summary}. 
The differences in these SM predictions are mainly due to the development of the $\Bb  \to D^{(\ast)}$ form factor evaluations both by theoretical studies and experimental fits.
For the moment, the HFLAV takes the ``arithmetic'' average~\cite{HFLAV2024winter} for the SM prediction based on the seven representative works~\cite{Bigi:2016mdz,Bernlochner:2017jka,Jaiswal:2017rve,Gambino:2019sif,Bordone:2019vic,BaBar:2019vpl,Martinelli:2021onb}, 
in which several form-factor parametrizations are used for their fit analyses. 
The important notices of these references are i) the CLN parameterization is no longer adopted and ii) the recent lattice studies~\cite{FermilabLattice:2021cdg,Aoki:2023qpa,Harrison:2023dzh} for $B \to D^*$ transition are not included.
In Table~\ref{tab:data_SM_summary}, we refer to Refs.~\cite{Bordone:2019vic,Bordone:2019guc} separately from the HFLAV average since it obtains the distinct $R_{D^*}^\text{SM}$. 
This study has considered the heavy quark effective theory (HQET) corrections up to $\mathcal{O}(\alpha_s, \Lambda_{\rm QCD}/m_b, \Lambda_{\rm QCD}^2/m_{c}^2)$,
and utilized the approximate $SU(3)_F$ relation to the $\Bb_s\to D_s^{(\ast)}$ form factors that have been determined by the HPQCD lattice result~\cite{McLean:2019qcx}.
Reference~\cite{Iguro:2020cpg} also takes the same HQET form factor as above but includes the $B \to D^{(*)}\ell\nu$ differential angular distribution data for the fit, which is distinct from the former study. 
For comparison, we also refer to the recent study of Ref.~\cite{Bernlochner:2022ywh} that has included the $B \to D^*$ lattice study by Fermilab--MILC (FM)~\cite{FermilabLattice:2021cdg}. 
In this work, we will employ the work of Ref.~\cite{Iguro:2020cpg} for the form factor evaluations as explained later.

A further concern for the SM evaluation is long-distance QED corrections to $\Bb \to D^{(\ast)}\ell\overline\nu$, which remains an open question. 
They depend on the lepton mass as being of $\mathcal{O}[\alpha \ln (m_\ell/m_B)]$ and hence it could provide a few percent corrections to violation of the LFU in the semileptonic processes\cite{deBoer:2018ipi,Cali:2019nwp,Isidori:2020eyd,Papucci:2021ztr}.
This will be crucial in the future when the LHCb and Belle~II experiments reach such an accuracy.

\begin{figure}[t]
\begin{center}
 \includegraphics[viewport=0 0 800 495, width=36em]{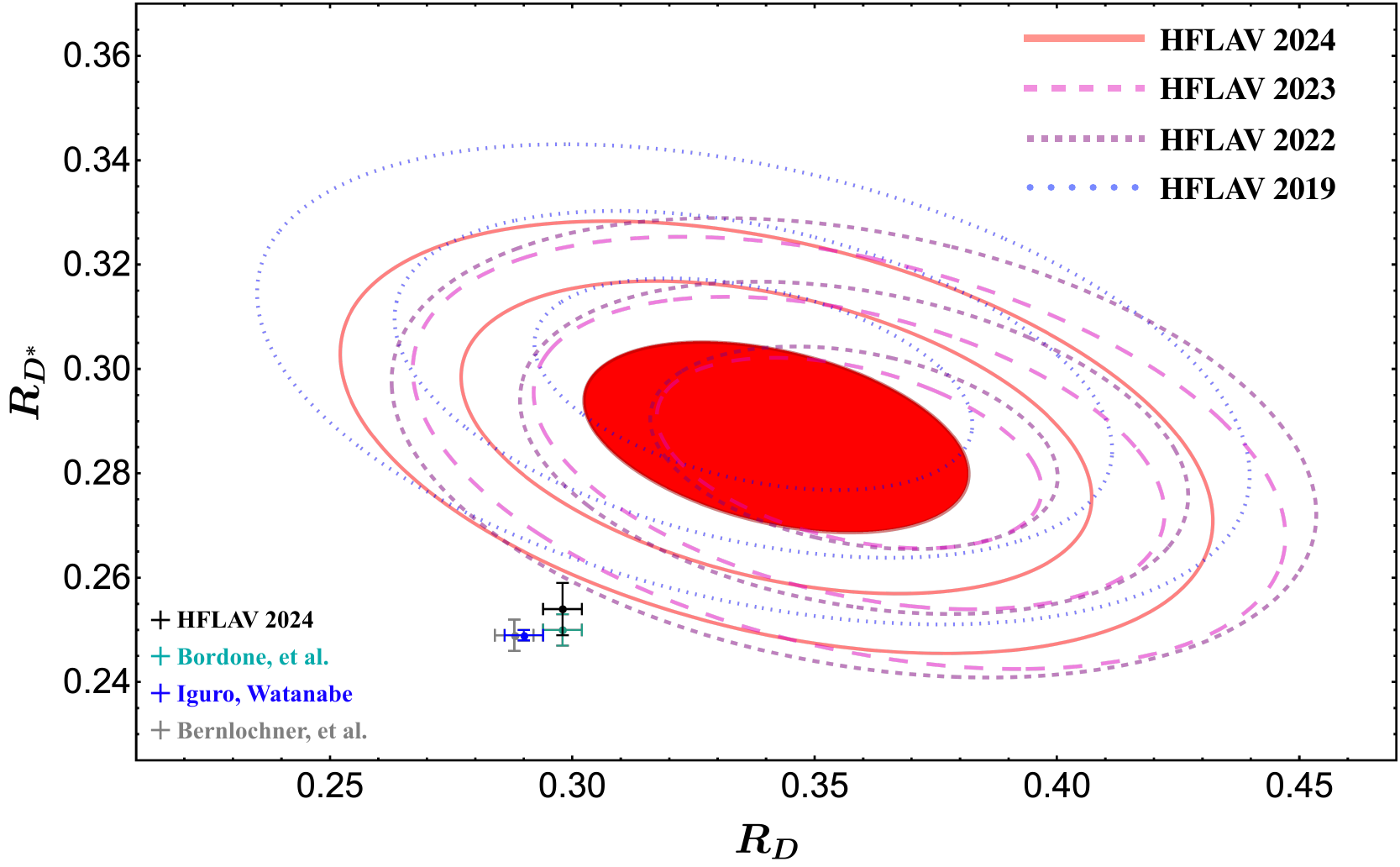}
\end{center}
 \caption{
 \label{fig:RDplot}
 The world average of the latest $R_D$ and $R_{D^*}$ experimental results by HFLAV 2024 (1,\,2,\,3$\sigma$ red-solid contours).
 The former world averages are also shown: the HFLAV 2023, 2022 and 2019 averages by long-dashed, dashed and dotted contours, respectively.
 On the other hand, the several SM predictions are shown by crosses \cite{HFLAV2024winter,Bernlochner:2022ywh,Iguro:2020cpg, Bordone:2019vic,Bordone:2019guc} (see also Table~\ref{tab:data_SM_summary}).
}
\end{figure}

In Fig.~\ref{fig:RDplot}, we show 
the latest average of the $R_D$--$R_{D^\ast}$ along with the several recent SM predictions. 
A general consensus from the figure is that the deviation of the experimental data from the SM expectations still remains.
For instance, applying the SM prediction from \{HFLAV2024~\cite{HFLAV2024winter}, Bernlochner~{\it et al.}~\cite{Bernlochner:2022ywh}, Iguro-Watanabe~\cite{Iguro:2020cpg}, Bordone~{\it et al.}~\cite{Bordone:2019vic,Bordone:2019guc}\}, 
one can see 
$\{3.3 \sigma, 4.2\sigma, 4.4 \sigma, 3.9 \sigma\}$ deviations corresponding to $p\text{-value}=\{8.9 \times 10^{-4}, 2.3 \times 10^{-5}, 1.1\times 10^{-5},  1.2 \times 10^{-4}\}$ ($\Delta \chi^2 = \{14.0, 21.4, 22.8, 18.1$\} for two degrees of freedom), respectively.\\

%%%Figure[begin]
%%%Figure[end]

In addition to these deviations in the LFU measurements, $\tau$- and $D^\ast$-polarization observables in $\Bb \to D^{(\ast)} \tau \overline{\nu}$ also provide us important and nontrivial information.
This is because these observables can potentially help us to pin down the NP structure that causes these deviations~\cite{Tanaka:2010se,Sakaki:2012ft,Duraisamy:2013pia,Duraisamy:2014sna,Becirevic:2016hea,Alok:2016qyh,Ivanov:2017mrj,Colangelo:2018cnj,Bhattacharya:2018kig,Iguro:2018vqb,Blanke:2018yud,Blanke:2019qrx,Becirevic:2019tpx,Hill:2019zja,Alguero:2020ukk,Bhattacharya:2020lfm,Penalva:2021gef,Penalva:2022vxy}.
We refer to the $\tau$ longitudinal-polarization asymmetry in $\Bb \to D^{(\ast)}\tau \overline{\nu}$ and the fraction of the $D^{\ast}$-longitudinal mode in $\Bb \to D^{\ast}\tau \overline{\nu}$ 
as $P_\tau^{D^{(\ast)}}$ and $F_{L}^{D^{\ast}}$, respectively. 
See Refs.~\cite{Tanaka:2012nw,Asadi:2018sym,Iguro:2018vqb} for their explicit definitions. 

In recent years, the first measurements for some of the above polarization observables have been reported by the Belle collaboration.  
It is summarized as $P_\tau^{D^\ast\,\text{Belle}} = -0.38 \pm 0.51_{\rm stat} \,^{+0.21}_{-0.16}{}_{\rm syst} $~\cite{Hirose:2016wfn} and $F_L^{D^\ast\,\text{Belle}} = 0.60 \pm 0.08_{\rm stat}  \pm 0.04_{\rm syst} $~\cite{Abdesselam:2019wbt}. 
Note that the latter is still a preliminary result.
More recently,  
the LHCb  collaboration reported the first preliminary result of $F_L^{D^\ast}$  using the combined dataset of the LHCb Run\,1 and  part of the Run\,2;
$F_L^{D^\ast,\,\text{LHCb}} = 0.43 \pm 0.06_{\rm stat} \pm 0.03_{\rm syst}$ \cite{LHCbFL,LHCb:2023ssl}.
Naively combining these results,
we obtain the average value as
\begin{align}
 F_L^{D^\ast} = 0.49 \pm 0.05\,,
 \end{align}
which is well consistent with the SM prediction,  
see Table~\ref{tab:data_SM_summary}.
Note that although $P_\tau^{D}$ is the most striking observable to disentangle the leptoquark (LQ) scenarios that can explain the anomalies, the experimental result does not exist so far.

Furthermore,
the $D^{\ast}$-longitudinal polarization in the electron and muon modes have been measured very precisely by the full Belle data set  $F_{L, {\rm Belle}}^{D^\ast\!, e} = 0.485 \pm  0.017_{\rm stat} \pm 0.005_{\rm sys}$ and 
$F_{L, {\rm Belle}}^{D^\ast\!, \mu} = 0.518 \pm  0.017_{\rm stat} \pm 0.005_{\rm sys}$ \cite{Belle:2023bwv}, and also the first Belle~II data 
$F_{L, {\rm BelleII}}^{D^\ast\!, e} = 0.520 \pm  0.005_{\rm stat} \pm 0.005_{\rm sys}$
and $F_{L, {\rm BelleII}}^{D^\ast\!, \mu} = 0.527 \pm  0.005_{\rm stat} \pm 0.005_{\rm sys}$ \cite{Belle-II:2023okj}.
We obtain the naive averaged value,
\begin{align}
\begin{aligned}
F_{L}^{D^\ast\!, e} & = 0.515 \pm 0.007\,, \\
F_{L}^{D^\ast\!, \mu} & =0.526 \pm 0.007\,.
\end{aligned}
\end{align}
They are also consistent with the SM predictions, $F_{L,\,\text{SM}}^{D^\ast\!, \ell} (\ell = e,\, \mu)=0.534 \pm 0.002$ \cite{Iguro:2020cpg,Fedele:2023ewe} (to be exact, there is a $2.7\sigma$ level tension in $F_{L}^{D^\ast\!, e}$).

Besides, the first direct measurement of the LFU test for the inclusive mode, $R_X \equiv \mathcal{B}(\Bb \to X \tau \overline{\nu}_{\tau})
/\mathcal{B}(\Bb \to X \ell \overline{\nu}_{\ell})$, has been performed by the Belle~II collaboration.  Here, $X$ indicates any hadronic final states coming from $b\to c l \overline\nu$ and $b\to u l \overline\nu$ processes.
A robust correlation is expected between $R_X$ and $R_{D^{(\ast)}}$ because the inclusive mode is dominated by the exclusive $D$ and $D^\ast$ modes.
Recently, the Belle~II collaboration  reported the preliminary result, 
$R_X = 0.228 \pm 0.016_\text{\rm stat} \pm 0.036_\text{\rm syst}$ \cite{BelleIIRX,Belle-II:2023aih}.
This result is not only consistent with the SM prediction 
$R_X^{\rm SM} =  0.223 \pm 0.005$ \cite{Freytsis:2015qca,Ligeti:2021six,Rahimi:2022vlv}, but also consistent with the $R_{D^{(\ast)}}$ anomalies \cite{Belle-II:2023aih}.
Note that based on the LEP data of $\mathcal{B}(b \to X \tau \bar\nu_\tau)$ \cite{ALEPH:2000vvi} and assuming that each individual $b$ hadron has the same width, 
$R_{X_c} = 0.223 \pm 0.030$ has been estimated 
\cite{Celis:2016azn,Kamali:2018bdp}, which is consistent with the Belle II result.
See Ref.~\cite{Kamali:2018bdp} for a detailed NP analysis of $R_{X_c}$.

%%%%%%%%%%%%%%%%%%%%%%%%%%%%%%%
\subsection{Preliminaries of our analysis}
%%%%%%%%%%%%%%%%%%%%%%%%%%%%%%%

The main points of this paper are that 
(i) we provide state-of-the-art numerical NP formulae for the observables relevant to the semi-tauonic $B$ decays and 
(ii) we revisit to perform global fits to the available $R_{D^{(\ast)}}$ measurements with respect to NP interpretations. 
It will be given by incorporating the following updates and concerns:
\begin{itemize}
 \item 
 Three new results of the LFU test from the LHCb collaboration ($R_{D^{(\ast)}}^{\rm LHCb2022}$, $R_{D^\ast}^{\rm LHCb2023}$, and $R_{D^{(\ast)}}^{\rm LHCb2024}$) 
 and the preliminary result from the Belle~II collaboration ($R_{D^\ast}^{\rm BelleII}$) are encoded in our global analysis as shown in Table~\ref{tab:RD_exps}.
\item 
 The preliminary result of the $D^\ast$-polarization fraction from the LHCb collaboration ($F_L^{D^\ast,\,\text{LHCb}}$) is encoded in our global analysis as shown in Table~\ref{tab:RD_exps}.
 \item 
 The recent development of the $\Bb\to D^{(\ast)}$ transition form factors is taken into account, which is summarized in the previous section.
 It is described by the HQET taking higher-order corrections up to $\mathcal{O}(\Lambda_{\rm QCD}^2/m_c^2)$ as introduced in Refs.~\cite{Bernlochner:2022ywh,Jung:2018lfu,Bordone:2019vic}. 
 We follow the result from the comprehensive theory$\,+\,$experiment fit analysis as obtained in Ref.~\cite{Iguro:2020cpg}.\footnote{
 To be precise, we employ the ``(2/1/0) fit'' result, preferred by their fit analysis.
 See the reference for details.} 
 \item
 A recent study introduced an approximation method to reduce independent parameters involving the $\mathcal{O}(\Lambda_{\rm QCD}^2/m_c^2)$ corrections in HQET \cite{Bernlochner:2022ywh}. 
 Although this affects some of the parameter fits for the form factors, 
 our reference values of $R_{D^{(\ast)}}^{\rm SM}$ from Ref.~\cite{Iguro:2020cpg} are consistent with those of Ref.~\cite{Bernlochner:2022ywh}, shown    in Table~\ref{tab:data_SM_summary}.
 Hence, we do not take this approximation in the present work. 
 \item
 The FM  Collaborations~\cite{FermilabLattice:2021cdg} presented the first lattice result of the form factors for $\Bb \to D^\ast \ell \overline \nu$ at the nonzero recoil points.
 The FM result with the light-lepton experimental data predicts $R_{D^\ast}^{\rm SM}  = 0.2484 \pm 0.0013$, which is well consistent with our reference value of Ref.~\cite{Iguro:2020cpg}. 
 Currently, JLQCD~\cite{Aoki:2023qpa} and HPQCD~\cite{Harrison:2023dzh} collaborations presented preliminary results at the nonzero recoil points.
 Since their results need to be finalized and compared with each other, 
 we do not include these lattice updates in the present work.
 See Refs.~\cite{Martinelli:2023fwm,Kapoor:2024ufg} for the recent theoretical studies based on these lattice results.
 \item 
 The dispersive matrix method can determine the form factors based on only the lattice data with the unitary bound, which is free from the parametrization method \cite{Martinelli:2021frl,Martinelli:2021onb,Martinelli:2021myh}.
 Even though the lattice uncertainties are still large, this method predicts $R_D^{\rm SM} =0.296\pm0.008$ \cite{Martinelli:2021myh} and $R_{D^*}^{\rm SM} =0.262\pm0.009$ \cite{Martinelli:2023fwm} (including above lattice results at the nonzero recoil), which are consistent with the world averages of $R_{D^{(\ast)}}$ in Eq.~\eqref{eq:WAvalues} and also the SM predictions of Ref.~\cite{Iguro:2020cpg} with less than $2\sigma$ level.
 However, it has been recently pointed out that
  this method provides an additional tension in $F_{L}^{D^\ast\!, \ell} (\ell = e,\, \mu)$ at $2.4\sigma$ level \cite{Fedele:2023ewe,Martinelli:2023fwm}.
  In the present work, we do not employ this method.
 \item
 Indirect LHC bounds from the high-$p_T$ mono-$\tau$ searches with large missing transverse energy~\cite{Greljo:2018tzh,Dumont:2016xpj,Altmannshofer:2017poe,Iguro:2017ysu,Abdullah:2018ets,Iguro:2018fni,Baker:2019sli,Marzocca:2020ueu,Iguro:2020keo,Endo:2021lhi,Jaffredo:2021ymt} 
 are concerned.  
 We impose the result of Ref.~\cite{Iguro:2020keo} that directly constrains the NP contributions to the $b\to c\tau\overline\nu$ current and accounts for the NP-scale dependence on the LHC bound, 
 which is not available in the effective-field-theory description.
 Requiring an additional $b$-tagged jet also helps to improve the sensitivity~\cite{Marzocca:2020ueu,Endo:2021lhi}.
 We will see how it affects the constraints in the LQ  scenarios.
 \item
 Similar sensitivity can be obtained by the $\tau^+\tau^-$ final state~\cite{ATLAS:2020zms,CMS:2022goy}.
 It is noted that about three standard deviations are reported by the CMS collaboration~\cite{CMS:2022goy}, which would imply the existence of leptoquark,
 while the ATLAS result~\cite{ATLAS:2020zms} has not found a similar excess.
 We need the larger statistics to confirm it, and thus, we do not include the constraint to be conservative.
\end{itemize}
In addition to the above points, we also investigate the following processes that are directly/indirectly related to the $b \to c\tau\overline\nu$ current:
\begin{itemize}
 \item 
 The LFU in $B_c\to \Jpsi\, l\,\overline\nu$ decays is connected to $R_{D^{(*)}}$ through the same $b \to c l \overline{\nu}$ currents.
 The LHCb collaboration has measured the ratio 
 $R_{\Jpsi}^{\rm LHCb} \equiv \mathcal{B}(B_c\to \Jpsi\,\tau\,\overline\nu)/\mathcal{B}(B_c\to \Jpsi\,\mu\,\overline\nu) = 0.71\pm 0.17\pm 0.18$~\cite{Aaij:2017tyk}.
 Furthermore, very recently, the CMS collaboration announced the first preliminary result of $R_{\Jpsi}$ using the $B$ parking data, 
 $R_{\Jpsi}^\text{CMS2023} =  0.17 {}^{+0.18}_{-0.17}{}_\text{stat}  {}^{+0.21}_{-0.22} {}_\text{syst} {}^{+0.19}_{-0.18}{}_\text{theory}$ 
 with the muonic $\tau$ tagging \cite{CMSRJpsi} and 
 $R_{\Jpsi}^\text{CMS2024} =  1.04 {}^{+0.50}_{-0.44}$
 with the hadronic $\tau$ tagging \cite{CMSRJpsi2}. 
 By naively averaging them, we obtain 
 \begin{align}
 R_{\Jpsi} = 0.61 \pm 0.18\,,
 \end{align}
 which is consistent with the SM prediction \cite{Harrison:2020nrv} at $1.9\sigma$ level, see Table~\ref{tab:data_SM_summary}.
 Although these results still have large experimental uncertainties,
 it would be useful in the future to test some NP scenarios for the sake of the $R_{D^{(*)}}$ anomalies. 
 We update the numerical formula for $R_{\Jpsi}$ in the presence of general NP contributions and put a prediction from our fit study. 
 \item 
 The $\Upsilon$ leptonic decays, $\Upsilon\to l^+ l^-$,  are potentially connected to $R_{D^{(\ast)}}$ once one specifies NP interactions to the bottom quark and leptons. 
 Although the SM contribution comes from a photon exchange, it is suppressed by the $\Upsilon$ mass squared.
 The sensitivity to NP is, therefore, not completely negligible,
 and the LFU of $R_{\Upsilon(nS)}\equiv  \mathcal{B}(\Upsilon(nS) \to \tau^+ \tau^-) / \mathcal{B}(\Upsilon(nS) \to \ell^+ \ell^-)$ 
 can be an important cross check of the $R_{D^{(*)}}$ anomalies.
 Furthermore, 
 the BaBar collaboration has reported 
 a result that slightly violates the LFU: $R_{\Upsilon(3S)}= 0.966 \pm 0.008 \pm 0.014$~\cite{Lees:2020kom}. 
 We investigate the theoretical correlations in several NP models.
\end{itemize}

This paper is organized as follows. 
In Sec.~\ref{sec_GF}, we put the numerical formulae for the relevant observables in terms of the effective Hamiltonian.
We also summarize the case for single-operator analysis. 
In Sec.~\ref{sec:Fit}, based on the generic study with renormalization-group running effects, 
we obtain relations among $R_{D}$, $R_{D^{\ast}}$, and $F_{L}^{D^{\ast}}$ in the LQ models and discuss their potential to explain the present data. 
Relations to the $\tau$ polarizations are also discussed.
In Sec.~\ref{sec:Upsi_decay}, we also investigate the LFU violation in the $\Upsilon$ decays and show its correlation with $b\to c \tau \overline \nu$ observables.
Finally, we conclude in Sec.~\ref{sec:conclusion} and correlations in the precision measurements for each NP scenario are summarized in Table~\ref{tab:summary_table}.

%%%%%%%%%%%%%%%%%%%%%%%%%%%%%%%
\section{General formulae for the observables}
\label{sec_GF}
%%%%%%%%%%%%%%%%%%%%%%%%%%%%%%%

At first, we describe general NP contributions to $b \to c\tau\overline\nu$ in terms of the effective Hamiltonian.
The operators relevant to the processes of interest are described as\footnote{The different naming schemes of the operators are often used \cite{Sakaki:2013bfa,Huang:2018nnq,Kou:2018nap}.
Our $C_{V_L},\,C_{V_R},\,C_{S_L}$, and $C_{S_R}$ correspond to $C_{V_1},\,C_{V_2},\,C_{S_2}$, and $C_{S_1}$, respectively.}
\begin{align}
 \label{eq:Hamiltonian}
 {\cal {H}}_{\rm{eff}}= 2 \sqrt2 G_FV_{cb}\biggl[ (1+C_{V_L})O_{V_L}+C_{V_R}O_{V_R}+C_{S_L}O_{S_L}+C_{S_
R}O_{S_R}+C_{T}O_{T}\biggl]\,,
\end{align}
with
\begin{align}
 &O_{V_L} = (\overline{c} \gamma^\mu P_Lb)(\overline{\tau} \gamma_\mu P_L \nu_{\tau})\,,& 
 &O_{V_R} = (\overline{c} \gamma^\mu P_Rb)(\overline{\tau} \gamma_\mu P_L \nu_{\tau})\,,&\nonumber \\
 &O_{S_L} = (\overline{c}  P_Lb)(\overline{\tau} P_L \nu_{\tau})\,,&
 &O_{S_R} = (\overline{c}  P_Rb)(\overline{\tau} P_L \nu_{\tau})\,,& \label{eq:operator} \\
 &O_{T} = (\overline{c}  \sigma^{\mu\nu}P_Lb)(\overline{\tau} \sigma_{\mu\nu} P_L \nu_{\tau}) \,,&
 \nonumber
\end{align}
where $P_L=(1-\gamma_5)/2$ and $P_R=(1+\gamma_5)/2$. 
The NP contribution is encoded in the Wilson coefficients (WCs) of $C_X$, normalized by the SM factor of $2 \sqrt2 G_FV_{cb}$. 
The SM corresponds to 
$C_{X} = 0$ for $X=V_{L,R}$, $S_{L,R}$, and $T$ 
in this description. 
We assume that the light neutrino is always left-handed and NP contributions are relevant to only the third-generation leptons ($\tau,\,\nu_\tau$).\footnote{%
See Refs. \cite{Iguro:2018qzf,Asadi:2018wea,Greljo:2018ogz,Robinson:2018gza,Babu:2018vrl,Mandal:2020htr,Penalva:2021wye} for models with the right-handed neutrino $\nu_R$.
It is noted that the $W^\prime$ is necessarily accompanied by $Z^\prime$ and thus the recent di-$\tau$ resonance search \cite{ATLAS:2020zms,CMS:2022goy} excludes the $W_R^\prime$ explanation \cite{IguroKEK}.}

Note that the leading $SU(2)_L \times U(1)_Y$ invariant operator, to generate the LFU violated type of the $O_{V_R}$ form, is given in dimension-eight 
as $(\overline{c}_R \gamma^\mu b_R)(\overline{L}^3 \gamma_\mu \tau^A L^3)(\widetilde H \tau^A H)$. 
This implies that $C_{V_R}$ in a NP model necessarily has an additional suppression
compared with the other operators generated from dimension-six operators. 
See Ref.~\cite{Asadi:2019zja} for a NP model that can generate the  $C_{V_R}$ contributions to $R_{D^{(\ast)}}$.

In the following parts, the observables for $\Bb \to D^{(\ast)} \tau \overline{\nu}$, $B_c \to \tau \overline{\nu}$, and $B_c\to \Jpsi\, \tau\overline{\nu}$ 
are evaluated with Eq.~\eqref{eq:Hamiltonian} at the scale $\mu = \mu_b = 4.18\,\text{GeV}$. 
The process $\Upsilon(nS) \to l^+ l^-$ will be described in detail in Sec.~\ref{sec:Upsi_decay}.

%%%%%%%%%%%%%%%%%%%%%%%%%%%%%%%
\subsection*{\underline{$\Bb \to D^{(\ast)} \tau \overline{\nu}$}}
%%%%%%%%%%%%%%%%%%%%%%%%%%%%%%%
In this work, we follow analytic forms of the differential decay rates for $\Bb \to D^{(\ast)} \tau \overline{\nu}$ obtained in Refs.~\cite{Sakaki:2013bfa,Sakaki:2014sea}. 
Regarding the form factors, we employ the general HQET-based description~\cite{Bordone:2019vic}, 
in which the heavy quark expansions~\cite{Caprini:1997mu,Bernlochner:2017jka} are taken up to 
NLO for $\epsilon_a = \alpha_s/\pi$, $\epsilon_b = \overline\Lambda/(2m_b)$ and NNLO for $\epsilon_c = \overline\Lambda/(2m_c)$ by recalling the fact $\epsilon_a \sim \epsilon_b \sim \epsilon_c^2$. 
Thanks to HQET property, the form factors for the different Lorenz structures of the NP operators are connected to that for the SM current, 
which enables us to evaluate the NP contributions to the observables.

Two parametrization models have been considered with respect to the $z = (\sqrt{w+1} -\sqrt 2) / (\sqrt{w+1} +\sqrt 2)$ expansions for the form factors in this description, 
with which the most general fit analyses of the form-factor parameters and $|V_{cb}|$ have been performed in Ref.~\cite{Iguro:2020cpg}. 
For the present work, we take the $(2/1/0)$ model with a minor update and apply the updated fit result based on Ref.~\cite{Iguro:2020cpg}.

We have evaluated the ratio observables, $R_{D^{(\ast)}}$, $P_\tau^{D^{(\ast)}}$ and $F_{L}^{D^{\ast}}$, 
for the case of the effective Hamiltonian of Eq.~\eqref{eq:Hamiltonian} at the scale $\mu = \mu_b$. 
In the end, we find the following updated numerical formulae,
\begin{align}
 \label{eq:RD}
 \frac{R_D}{R_{D}^\textrm{SM}} =
 & ~|1+C_{V_L}+C_{V_R}|^2  + 1.01|C_{S_L}+C_{S_R}|^2 + 0.84|C_{T}|^2  \nonumber \\[-0.5em]
 & + 1.49\textrm{Re}[(1+C_{V_L}+C_{V_R})(C_{S_L}^*+C_{S_R}^*)]  + 1.08\textrm{Re}[(1+C_{V_L}+C_{V_R})C_{T}^*] \,, 
 \\[1em]
%\end{align}
 %
%\begin{align}
 \label{eq:RDs}
 \frac{ R_{D^{\ast}}}{R_{D^\ast}^\textrm{SM}} =
 & ~|1+C_{V_L}|^2 + |C_{V_R}|^2  + 0.04|C_{S_L}-C_{S_R}|^2 + 16.0|C_{T}|^2 \nonumber \\[-0.5em]
 & -1.83\textrm{Re}[(1+C_{V_L})C_{V_R}^*]  - 0.11\textrm{Re}[(1+C_{V_L}-C_{V_R})(C_{S_L}^*-C_{S_R}^*)] \nonumber \\%[0.5em] 
 & -5.17\textrm{Re}[(1+C_{V_L})C_{T}^*] + 6.60\textrm{Re}[C_{V_R}C_{T}^*] \,, 
 \\[1em]
%\end{align}
%
%\begin{align}
 \label{eq:PtauD}
 \frac{P_\tau^{D}}{P_{\tau,\,\textrm{SM}}^{D}} = 
 & \left(\frac{R_D} {R_D^\text{SM}}\right)^{-1} \!\!\!\times \Big( |1+C_{V_L}+C_{V_R}|^2  + 3.04|C_{S_L}+C_{S_R}|^2 + 0.17|C_{T}|^2 \nonumber \\
 & + 4.50\textrm{Re}[(1+C_{V_L}+C_{V_R})(C_{S_L}^*+C_{S_R}^*)]  -1.09\textrm{Re}[(1+C_{V_L}+C_{V_R})C_{T}^*] \Big)  \,, 
 \\[1em]
 \label{eq:PtauDs}
 \frac{P_\tau^{D^{\ast}}} {P_{\tau,\,\textrm{SM}}^{D^{\ast}}} = 
 & \left(\frac{R_{D^{\ast}}}  {R_{D^{\ast}}^\text{SM}}\right)^{-1} \!\!\!\times \Big( |1+C_{V_L}|^2  + |C_{V_R}|^2  - 0.07|C_{S_R}- C_{S_L}|^2 - 1.85 |C_{T}|^2 \nonumber \\
 & 
 - 1.79 \textrm{Re}[(1+C_{V_L})C_{V_R}^{\ast}]  
 + 0.23 \textrm{Re}[(1+C_{V_L}- C_{V_R})(C_{S_L}^* -C_{S_R}^*)]  \nonumber \\[0.5em]
 & 
 - 3.47 \textrm{Re}[(1+C_{V_L}) C_{T}^*]
 + 4.41  \textrm{Re}[C_{V_R} C_{T}^*] \Big)\,, 
% \\[1em]
\end{align}
\begin{align}
 \label{eq:FLDs}
 \frac{F_L^{D^*}}{F_{L,\,\textrm{SM}}^{D^{\ast}}} = 
 & \left(\frac{R_{D^*}}{R_{D^*}^\text{SM}}\right)^{-1} \!\!\!\times \Big( |1+C_{V_L}-C_{V_R}|^2  + 0.08|C_{S_L}-C_{S_R}|^2 + 6.90|C_{T}|^2 \nonumber \\
 & - 0.25\textrm{Re}[(1+C_{V_L}-C_{V_R})(C_{S_L}^*-C_{S_R}^*)]  -4.30\textrm{Re}[(1+C_{V_L}-C_{V_R})C_{T}^*] \Big) \,,
\end{align}
which can be compared with those in the  literature~\cite{Feruglio:2018fxo,Asadi:2018wea,Blanke:2018yud,Iguro:2018vqb,Mandal:2020htr}. 
The SM predictions are obtained as\footnote{
We updated the fit analysis with the modification of the formula for unitarity bound~\cite{Caprini:1997mu}, pointed out by Ref.~\cite{Bigi:2017jbd}. 
It only affects the last digits of the SM predictions, though.
} 
\begin{align}
    \begin{aligned}
   R_{D}^\textrm{SM} &= 0.290 \pm 0.003\,,\\[0.5em]
   R_{D^\ast}^\textrm{SM}& = 0.248\pm 0.001\,,\\[0.5em]
   P_{\tau,\,\textrm{SM}}^{D} &= 0.331 \pm 0.004\,,\\[0.5em]
   P_{\tau,\,\textrm{SM}}^{D^{\ast}} &= -0.497 \pm 0.007\,,\\[0.5em]
   F_{L,\,\textrm{SM}}^{D^{\ast}} &= 0.464 \pm 0.003\,.
    \end{aligned}
\end{align}

Furthermore, we have checked uncertainties of the above numerical coefficients in the formulae,
based on the fit result from Ref.~\cite{Iguro:2020cpg}. 
The tensor (scalar) terms involve $\sim4\%\, (10\%)$ uncertainties for the $D$ ($D^*$) mode,  while the others contain less than $1\%$ errors. 
At present, they are not significant and thus neglected in our following study.

%%%%%%%%%%%%%%%%%%%%%%%%%%%%%%%
\subsection*{\underline{$B_c \to \tau \overline{\nu}$}}
%%%%%%%%%%%%%%%%%%%%%%%%%%%%%%%
The significant constraint 
on the scalar operators $O_{S_{L,R}}$ 
comes from the $B_c$ lifetime measurements ($\tau_{B_c}$) \cite{Beneke:1996xe,Alonso:2016oyd,Celis:2016azn,Watanabe:2017mip,Aebischer:2021ilm}:
the branching ratio of $B_c^- \to \tau \overline\nu$, which has not yet been observed, 
is significantly amplified by the NP scalar interactions, and the branching ratio is constrained from measured $\tau_{B_c}$ \cite{Zyla:2020zbs}.
We obtain an upper bound on the WCs as
\begin{align}
 \label{eq:Bc}
 \left| 1 + C_{V_L} - C_{V_R} - 4.35\, (C_{S_L} - C_{S_R}) \right|^2 = \frac{\mathcal{B}(B_c\to \tau\overline\nu)}{\mathcal{B}(B_c\to \tau\overline\nu)_{\rm SM}} < 
27.1\left( \frac{\mathcal{B}(B_c\to \tau\overline\nu)_{\rm UB}}{0.6}\right)\,,
\end{align}
with $\mathcal{B}(B_c\to \tau\overline\nu)_{\rm SM} \simeq 0.022$.
Throughout this paper, $|V_{cb}| = \left(41.0\pm 1.4\right) \times 10^{-3}$ is used  unless otherwise mentioned \cite{Zyla:2020zbs}.
The $b$ and $c$ quark mass inputs, which are relevant for scalar contributions, are taken as $m_b(\mu_b)=(4.18 \pm 0.03)\,\text{GeV}$ and $m_c(\mu_b)=(0.92\pm0.02)\,\text{GeV}$. 
Reference \cite{Alonso:2016oyd} evaluated that 
the upper bound (UB) on the branching ratio from $\tau_{B_c}$ is 
$\mathcal{B}(B_c\to \tau\overline\nu)_{\rm UB}=0.3$.
However, it is pointed out by Ref.~\cite{Blanke:2018yud} and later confirmed by Ref.~\cite{Aebischer:2021ilm} 
that there is a sizeable charm-mass dependence on the $B_c$ decay rate because the dominant contribution comes from the charm-quark decay into strange within the $B_c$ meson.
A conservative bound is set by Ref.~\cite{Blanke:2018yud}
as $\mathcal{B}(B_c\to \tau\overline\nu)_{\rm UB}=0.6$.

One should note that 
more aggressive bound $\mathcal{B}(B_c \to \tau \overline{\nu})_{\rm UB} = 0.1$ has been obtained in Ref.~\cite{Akeroyd:2017mhr} by using LEP data.
However, it is pointed out that $p_T$ dependence 
of the fragmentation function, $b \to B_c$, has been entirely overlooked, 
and thus the bound must be overestimated by several factors \cite{Blanke:2018yud,Bardhan:2019ljo,Blanke:2019qrx}. 
Although the CEPC and FCC-ee experiments
are in the planning stages,
the future Tera-$Z$ machines can directly measure $\mathcal{B}(B_c\to\tau\overline\nu)$ 
at O(1$\%$) level \cite{Zheng:2021xuq,Amhis:2021cfy,Zuo:2023dzn}.

Thanks to the conservative bound, the left-handed scalar operator, $C_{S_L}$ comes back to the game.
For instance, a general two-Higgs doublet model is a viable candidate, and readers are referred to Refs.\cite{Iguro:2022uzz,Blanke:2022pjy}.

%%%%%%%%%%%%%%%%%%%%%%%%%%%%%%%
\subsection*{\underline{$B_c\to \Jpsi\, \tau\,\overline{\nu}$}}
%%%%%%%%%%%%%%%%%%%%%%%%%%%%%%%
We follow the form factor description from the recent lattice result of Ref.~\cite{Harrison:2020gvo} for $B_c\to \Jpsi\, \tau\,\overline{\nu}$. 
We also take $m_b(\mu_b)$ and $m_c(\mu_b)$ for the scalar and tensor sectors as aforementioned. 
The formula for $R_{\Jpsi}$ is given as 
\begin{align}
\begin{aligned}
 \frac{ R_{\Jpsi}}{R_{\Jpsi}^\textrm{SM}} =
 & ~|1+C_{V_L}|^2 + |C_{V_R}|^2  +0.04|C_{S_L}-C_{S_R}|^2 + 14.7|C_{T}|^2  \\
 & -1.82\textrm{Re}[(1+C_{V_L})C_{V_R}^*]  - 0.10\textrm{Re}[(1+C_{V_L}-C_{V_R})(C_{S_L}^*-C_{S_R}^*)]  \\[0.5em] 
 & -5.39\textrm{Re}[(1+C_{V_L})C_{T}^*] + 6.57\textrm{Re}[C_{V_R}C_{T}^*] \,, 
 \end{aligned}
\end{align}
where we take $R_{\Jpsi}^\textrm{SM} = 0.258 \pm 0.004$~\cite{Harrison:2020nrv}. 
The coefficients potentially have $10$--$20\%$ uncertainties for $C_{S_{L,R}}$ and $C_T$, while a few percent for $C_{V_{L,R}}$. 

It is also known that there is a good NP correlation in the general effective Hamiltonian  \cite{Watanabe:2017mip,Yasmeen:2024cki}
\begin{align}
\frac{ R_{\Jpsi}}{R_{\Jpsi}^\textrm{SM}} \simeq \frac{R_{D^\ast}}{R_{D^\ast}^{\rm SM}}\,. \label{eq:JpsiDst}
\end{align}
This is because both channels are the scalar to vector-meson transitions. 
Taking the aforementioned averages of $R_{\Jpsi}^\text{exp}$ and $R_{D^*}^\text{exp}$, we see
\begin{align}
\frac{ R_{\Jpsi}^\text{exp}}{R_{\Jpsi}^\textrm{SM}} - \frac{R_{D^\ast}^\text{exp}}{R_{D^\ast}^{\rm SM}} = 1.2 \pm 0.7\,, 
\end{align}
which shows $1.7\sigma$ deviations from the prediction of Eq.~\eqref{eq:JpsiDst} for now but it is not conclusive due to large uncertainty from $R_{\Jpsi}^\text{exp}$.

%%%%%%%%%%%%%%%%%%%%%%%%%%%%%%%
\subsection*{\underline{$\Lambda_b \to \Lambda_c \tau \overline{\nu}$}}
%%%%%%%%%%%%%%%%%%%%%%%%%%%%%%%

A baryonic counterpart of the $b \to c \tau \overline\nu$ decay is $\Lambda_b \to \Lambda_c \tau \overline\nu$.
 Normalizing by the light-lepton channels,
the LFU observable $R_{\Lambda_c}$ is defined, 
$
R_{\Lambda_c} \equiv \mathcal{B}(\Lambda_b \to \Lambda_c \tau \overline \nu)/\mathcal{B}(\Lambda_b \to \Lambda_c \ell \overline \nu)$.
Similar to the $R_{D^{(\ast)}}$, 
the CKM dependence completely drops out 
and the form-factor uncertainties  are significantly reduced in $R_{\Lambda_c}$~\cite{Gutsche:2015mxa,Shivashankara:2015cta,Detmold:2015aaa,Li:2016pdv,Datta:2017aue,Bernlochner:2018kxh,Bernlochner:2018bfn}.
Furthermore, since there is no subleading Isgur-Wise function at $\mathcal{O}(\bar \Lambda/m_{c,b})$
in the $\Lambda_b \to \Lambda_c$ transition, the theoretical uncertainty is even suppressed~\cite{Neubert:1993mb}.
As one can easily imagine, $R_{D^{(\ast)}}$ and $R_{\Lambda_c}$ have a strong theoretical correlation through the $b \to c \tau \overline\nu$ interaction.
NP contributions with these correlations have been 
widely investigated including the forward-backward asymmetry of $\tau$, the longitudinal polarizations of $\Lambda_c$ and $\tau$, and a leptonic energy distribution 
\cite{DiSalvo:2018ngq,Hu:2018veh,Blanke:2018yud,Ray:2018hrx,Murgui:2019czp,Blanke:2019qrx,Penalva:2019rgt,Ferrillo:2019owd,Mu:2019bin}.

Based on the lattice QCD results for the $\Lambda_b \to \Lambda_c$ transition \cite{Detmold:2015aaa, Datta:2017aue,Murgui:2019czp}, 
we obtain a numerical formula of $R_{\Lambda_c}$ \cite{Fedele:2022iib} 
\beq
\label{eq:RLambda}
\frac{R_{\Lambda_c}}{R_{\Lambda_c}^{\rm SM}}& =  |1+C_{V_L}|^2
+ |C_{V_R}|^2 
- 0.72\, \textrm{Re}[ (1 +C_{V_L}) C_{V_R}^\ast]
+ 0.50 \,\textrm{Re}[ (1 +C_{V_L}) C_{S_R}^\ast +
C_{V_R} C_{S_L}^\ast]
\nonumber \\
& \quad + 0.33 \,\textrm{Re}[ (1 +C_{V_L}) C_{S_L}^\ast 
+ 
C_{V_R} C_{S_R}^\ast]
+ 0.52 \, \textrm{Re}[C_{S_L} C_{S_R}^\ast ]
+ 0.32 \, (|C_{S_L}|^2 + |C_{S_R}|^2) 
\nonumber \\[0.5em]
&\quad  -3.11 \,\textrm{Re}[ (1 +C_{V_L})  C_T^\ast]
+ 4.88\, \textrm{Re}[ C_{V_R}  C_T^\ast]
+ 10.4 \, |C_T|^2\,,
\eeq
where we again take the same $m_b(\mu_b)$ and $m_c(\mu_b)$ as above for the form factors of the scalar and pseudoscalar currents.
The SM prediction is 
$R_{\Lambda_c}^{\rm SM} = 0.324 \pm 0.004$ \cite{Bernlochner:2018kxh,Bernlochner:2018bfn}, 
where the LHCb data of $d \Gamma (\Lambda_b \to \Lambda_c \mu \overline \nu)/ d q^2$ \cite{Aaij:2017svr} is used for a fit of their HQET parameters,  in addition to the lattice QCD  form factor calculations.

It is known that 
there is a sum rule between 
$R_{\Lambda_c}$ and $R_{D^{(\ast)}}$
 \cite{Blanke:2018yud,Blanke:2019qrx,Fedele:2022iib}
\beq
\frac{R_{\Lambda_c}}{R_{\Lambda_c}^{\rm SM}}=
0.280\, \frac{R_{D}}{R_{D}^{\rm SM}} + 0.720\,\frac{R_{D^\ast}}{R_{D^\ast}^{\rm SM}} + \delta_{\Lambda_c}\,,
\eeq
with
\beq
\delta_{\Lambda_c} =&\,  
\textrm{Re}\left[(1 +C_{V_L} ) 
( 
0.035\,  C_{V_R}^\ast
- 0.003\,   C_{S_R}^\ast
 + 0.314\, C_T^\ast 
)\right]
- \textrm{Re}[C_{V_R} (0.003 \,C_{S_L}^\ast  + 0.175\, C_T^\ast )]
\nonumber \\
& +0.014\,( |C_{S_L}|^2 + |C_{S_R}|^2)
- 1.30\,|C_T|^2
+ 0.004 \,\textrm{Re}[ C_{S_L} C_{S_R}^\ast ] \,.
\eeq
As far as $|C_T| \ll 1$ holds, 
 $\delta_{\Lambda_c}$ never be relevant and this sum rule holds in any NP scenario.
Ignoring the small $\delta_{\Lambda_c}$  term,
we obtain a model-independent prediction of $R_{\Lambda_c}$
\beq
R_{\Lambda_c} &\simeq R_{\Lambda_c}^{\rm SM}
\left(
0.280\, \frac{R_{D}}{R_{D}^{\rm SM}} + 0.720\,\frac{R_{D^\ast}}{R_{D^\ast}^{\rm SM}}
\right)\nonumber \\
 & = R_{\Lambda_c}^{\rm SM} \left( 1.163 \pm 0.034 \right)  \nonumber \\[0.45em]
 &= 0.377 \pm 0.011_{R_{D^{(\ast)}}} \pm 0.005_{R_{\Lambda_c}^{\rm SM}}\,.
\label{eq:sum}
\eeq
This implies that $R_{\Lambda_c}$ is not used to distinguish the NP scenarios but rather gives a consistency check among the experimental measurements.
While the $\Lambda_b \to \Lambda_c \ell \overline\nu$ decay 
has been experimentally measured already with good accuracy \cite{Abdallah:2003gn,Aaltonen:2008eu,Aaij:2015bfa,Aaij:2017svr},
the $\Lambda_b \to \Lambda_c \tau \overline\nu$
decay had not been observed until 2022.
The observed value of the LHCb collaboration is
$R_{\Lambda_c}^{\rm LHCb} = 0.242 \pm  0.026_{\rm stat} \pm 0.071_{\rm syst}$
\cite{LHCb:2022piu}, which provides 
 a $1.8\sigma$ level tension from the sum rule prediction in Eq.~\eqref{eq:sum}.
 Instead, normalizing with the SM prediction of $\Gamma(\Lambda_b\to\Lambda_c\mu\overline{\nu})$ improves the accuracy and slightly up-lifts the central value, 
$R_{\Lambda_c} = |0.041/V_{cb}|^2 (0.271 \pm 0.069) = 0.271 \pm 0.072$ \cite{Bernlochner:2022hyz}.
While suppression of $R_{\Lambda_c}$ compared to the sum rule prediction would not be compatible with NP scenarios for $R_{D^{(\ast)}}$ anomaly, the experimental uncertainty in $R_{\Lambda_c}$ is still very large to draw a clear-cut conclusion.
Detailed analysis for NP scenarios for light lepton modes is given in Ref.~\cite{Fedele:2022iib}.

%%%%%%%%%%%%%%%%%%%%%%%%%%%%%%%
\section{Fit analysis}
\label{sec:Fit}
%%%%%%%%%%%%%%%%%%%%%%%%%%%%%%%
In this paper, 
we perform the following statistical analysis to probe several NP scenarios via a bottom-up approach:
\begin{enumerate}
 \item 
 Three measurements of $R_{D}$,  $R_{D^{*}}$  and $F_L^{D^*}$ are taken in the $\chi^2$ fit, 
 and then the favored regions for the NP parameter space are obtained, which are defined by $\Delta\chi^2 = \chi^2 - \chi^2_\text{NP-best} \leq 1$.
 \item 
 We then check whether the above solutions are consistent with the other relevant observables, such as the $B_c$ lifetime and the LHC bounds. 
 \item
 Furthermore, we evaluate NP predictions on $P_\tau^{D}$, $P_\tau^{D^{*}}$ and $R_{\Jpsi}$, where the above step 1 and 2 are passed.
 \item
 If applicable, a combined study with $R_{\Upsilon(3S)}$ is discussed. 
\end{enumerate}
The $\chi^2$ fit function in the step 1 is defined as 
\begin{align}
    \chi^2\equiv\sum_{i,j} (O^{\rm{theory}} 
-O^{\rm{exp}})_i ~{\rm Cov}^{-1}_{ij} 
~(O^{\rm{theory}}-O^{\rm exp})_{j} \,, 
\end{align}
where we take into account the $R_{D^{(*)}}$ and $F_L^{D^*}$ measurements for $O^{\rm{exp}}$ summarized by HFLAV as shown in Sec.~\ref{sec:intro}. The covariance is given as ${\rm Cov}_{ij} = \Delta O^{\rm{exp}}_i \rho_{ij} \Delta O^{\rm{exp}}_j + \Delta O^{\rm{theory}}_i \delta_{ij} \Delta O^{\rm{theory}}_j$, 
where correlation $\rho_{ij}$ is given as in Table~\ref{tab:RD_exps} while $\rho_{ij} = \delta_{ij}$ among the independent measurements. 
For every observable, we have the theory formulae $O^{\rm{theory}}$ as shown in Sec.~\ref{sec_GF}, 
and hence obtain best-fit values in terms of the WCs as defined in Eq.~\eqref{eq:Hamiltonian}, which are given at the $\mu_b$ scale.

Given the SM predictions as $R_{D}^\textrm{SM} = 0.290 \pm 0.003$, $R_{D^\ast}^\textrm{SM} = 0.248 \pm 0.001$, and $F_{L,\,\textrm{SM}}^{D^{\ast}} = 0.464 \pm 0.003$ \cite{Iguro:2020cpg}, 
we obtain $\chi^2_\text{SM} = 23.1$ (corresponding to $4.1\sigma$) implying a large deviation from the SM. 
Briefly, $\chi_{\text{NP-best}}^2 \leq  \mathcal{O}(1)$ implies an excellent fit by the NP operators, but its criterion depends on the number of the fitted WCs.
In our analysis,  the goodness-of-fit for each NP scenario (the likelihood-ratio test between the best-fit points) is expressed by the ``Pull'' value (defined in, \eg, Refs.~\cite{Descotes-Genon:2015uva,Blanke:2018yud}).
It depends on the number of the fitted WCs.
For cases of the single WC fits, the Pull is equivalent to
\begin{align}
 \text{Pull}\, (\text{single~WC}) =  \sqrt{\chi^2_\text{SM} - \chi^2_\text{NP-best}} ~(\sigma) \,. 
\end{align}
Therefore, Pull $\gtrsim 4$ represents an excellent NP fit, where 
$R_D$ and $R_{D^{*}}$ anomalies mostly can be explained at $1\sigma$ levels. Thus, 
we can quantitatively compare the NP scenarios by using the Pull values.

%%%Table[begin]
%%%%%%%%%%%%%%%%%%%%%%%%%%%%%%%
\begin{table}[t]
\centering
\newcommand{\bhline}[1]{\noalign{\hrule height #1}}
\renewcommand{\arraystretch}{1.5}
\rowcolors{2}{gray!15}{white}
\addtolength{\tabcolsep}{5pt} % add space between columns
 \begin{adjustbox}{width=\columnwidth,center}
  \begin{tabular}{cccccc} 
  %\toprule
  \bhline{1 pt}
  \rowcolor{white}
  &$|C_{V_L}|$ & $|C_{V_R}|$ & $|C_{S_L}|$ & $|C_{S_R}|$ & $|C_{T}|$  \\    
  \hline
 EFT ($>10\,\text{TeV}$) & $  0.32~(0.09) $ & $  0.33~(0.09) $ &$   0.55~(0.14) $ & $ 0.55~(0.15) $ &  $  0.17~(0.04)$ \\ %\hline
LQ ($4\,\text{TeV}$) & $ 0.36~(0.10)  $ & $0.40~(0.10)$ & $0.74~(0.17)$ & $0.67~(0.18)$  & $0.22~(0.05)$  \\ %\hline
LQ ($2\,\text{TeV}$) & $  0.42~(0.12) $ & $ 0.51~(0.15) $ & $ 0.80~(0.22) $ & $ 0.77~(0.22) $ & $ 0.30~(0.07)$ \\ 
\bhline{1 pt}
%\bottomrule
 \end{tabular}
 \end{adjustbox}
\addtolength{\tabcolsep}{-5pt} % set back to normal
    \caption{\label{tab:LHC_bound_one}
The $95\%$ CL upper bounds 
on the WCs at the $\mu = \mu_b$ scale from the LHC analysis of the $\tau$\,$+$\,missing search \cite{Iguro:2020keo}. 
The future prospects with $b$-tagged jet\,$+\,\tau \nu$ final state assuming 3\,ab$^{-1}$ of accumulated data are given in the parenthesis \cite{Endo:2021lhi}.
The NP mass scale is shown as $M_{\rm LQ} = 2$\,TeV, $4\,$TeV and $\Lambda_{\rm EFT} >10\,$TeV.
}
\end{table}
%%%%%%%%%%%%%%%%%%%%%%%%%%%%%%%
%%%%%%%%%%%%%%%%%%%%%%%%%%%%%%%
%%%Table[end]
Regarding the LHC bound to be compared with the above fit result, we refer to the result from Ref.~\cite{Iguro:2020keo}, in which the $\tau + \text{missing}$ searches have been analyzed. 
Their result is shown in Table~\ref{tab:LHC_bound_one}, where we give the $95\%$ CL upper limit at the $\mu_b$ scale.\footnote{
Note that Table~2 of Ref.~\cite{Iguro:2020keo} shows the
LHC bound at $\mu = \Lambda_\text{LHC}$. 
}
It should be emphasized that the LHC bound on the WC has a non-negligible mediator mass dependence, see Ref.~\cite{Iguro:2020keo} for details. 
This feature is indeed crucial for some NP scenarios, as will be seen later.
Furthermore, it is pointed out that the charge asymmetry of the $\tau$ lepton will improve the bound on $C_X$.

%%%%%%%%%%%%%%%%%%%%%%%%%%%%%%%
\subsection{EFT: single operator scenario}
\label{sec:singleWCfit}
%%%%%%%%%%%%%%%%%%%%%%%%%%%%%%%
We begin with the single NP operator scenarios based on the effective field theory (EFT) of Eq.~\eqref{eq:Hamiltonian}.
Assuming the WC to be real, we immediately obtain the fit results with the Pull values
and predictions of $P_\tau^D,~P_{\tau}^{D^\ast}$ and $R_{\Jpsi}$ as shown in Table~\ref{tab:fit:singlereal}. 
The allowed regions from the $B_c$ lifetime and current LHC bounds are listed as well.

For all the NP scenarios, we can see much improvement in the fit compared with the SM.
A significant change from the previous conclusion (before the new LHCb results~\cite{LHCbRun1,LHCb:2023zxo,LHCbRun2,LHCb:2023uiv}  came up~\cite{Blanke:2019qrx,Murgui:2019czp,Shi:2019gxi}) is that 
the $C_{S_R}$ scenario {\em becomes consistent with the data within 95\% CL}, \ie, $\chi^2_{best} < 8.0$ (for three observed data).
Unfortunately, it is known that the usual type-II  two-Higgs doublet model (2HDM) cannot achieve this solution because
the sign of $C_{S_R}$ must be negative: $C_{S_R} = - m_bm_\tau \tan^2\beta/m_{H^\pm}^2 <0$.
It is noted that even in the generic 2HDM, sizable $C_{S_R}$ contribution is difficult due to constraints from $\Delta M_{s}$ and the LHC search~\cite{Faroughy:2016osc,Iguro:2017ysu}.
Instead, the real $C_{S_L}$ scenario is not likely to explain the present data within $2\sigma$ level.
This is mainly because that the coefficients of interference terms between the real $C_{S_L}$ and the SM contribution in Eqs.~\eqref{eq:RD} and \eqref{eq:RDs}
have opposite signs, 
which prevent the simultaneous explanation of $R_{D^{(\ast)}}$ anomalies.
The situation is the same as the previous one before the new LHCb and Belle~II data.
The $C_{V_L}$ scenario well explains the present data, while $C_{V_R}$ gives a lower Pull.
The $C_T$ solution gives unique predictions on the other observables, which may be able to identify the NP scenario, and
it predicts a negative shift of $F_L^{D^\ast}$ with a tension from the present measurement~\cite{Abdesselam:2019wbt,Iguro:2018vqb}.\\ 

%%%%%%%%%%%%%%%%%%%%%%%%%%%%%%%
%%%Table[begin]
\begin{table}[t]
\centering
\newcommand{\bhline}[1]{\noalign{\hrule height #1}}
\renewcommand{\arraystretch}{1.5}
\rowcolors{3}{white}{gray!15}
%\addtolength{\tabcolsep}{5pt} % add space between columns
  % \scalebox{0.8}{
   \begin{adjustbox}{width=\columnwidth,center}
  \begin{tabular}{l c c|c c|c c c} 
  \bhline{1 pt}
 \multirow{2}{*}{ } & 
 \multirow{2}{*}{Pull~[$\chi^2_\text{best}$]} & 
 \multirow{2}{*}{Fitted $C_X$} &
 \multicolumn{2}{c|}{Allowed region of $C_X$}&
 \multicolumn{3}{c}{Predictions $(\Delta\chi^2 \le 1)$} \\
 %\hline
 %
 & & %$(\Delta \chi^2 \le 1)$ 
 & $B_c \to \tau \overline \nu$ & LHC & $P_\tau^D$ & $-P_\tau^{D^*}$ & $R_{\Jpsi}$  \\
 \hline
 SM &
 --${~[23.1]}$ &
 -- &
 -- &  -- &
 $0.331 \pm 0.004$ &
 $0.497 \pm 0.007$ &
 $0.258 \pm 0.004$ \\
 \hline
$C_{V_L}$ &
 ${4.8~[0.3]}$ & %updated-17-July-2024
 ${+0.079(16)}$ & %updated-17-July-2024
 very loose &  $[-0.32,0.32]$ &
 ${[0.331, 0.331]}$ & %2023 ver check
 ${[0.497, 0.497]}$ & %2023 ver check
 ${[0.291, 0.309]}$ \\ %updated-17-July-2024
 %\hline
 %
 %
 $C_{V_R}$ &
 ${2.6~[16.3]}$ & %updated-17-July-2024
 ${-0.070(26)}$ & %updated-17-July-2024
 very loose &  $[-0.33,0.33]$ &
 ${[0.331, 0.331]}$ & %2023 ver check
 ${[0.495, 0.496]}$ & %2023 ver check
 ${[0.280, 0.307]}$ \\ %updated-17-July-2024 
 %\hline
 %
 %
 $C_{S_L}$ &
 ${3.1~[13.6]}$ & %updated-17-July-2024
 ${+0.165(48)}$ & %updated-17-July-2024
 $[-0.94,1.4]$ &  $[-0.55,0.55]$ &
 ${[0.435, 0.508]}$ & %updated-17-July-2024
 ${[0.516, 0.531]}$ & %updated-17-July-2024
 ${[0.254, 0.256]}$ \\ %updated-17-July-2024
 %\hline
 %
 %
 $C_{S_R}$ &
 ${3.9~[8.0]}$ & %updated-17-July-2024
 ${+0.182(42)}$ & %updated-17-July-2024
 $[-1.4,0.94]$ &  $[-0.55,0.55]$ &
 ${[0.454, 0.516]}$ & %updated-17-July-2024 
 ${[0.458, 0.473]}$ & %updated-17-July-2024
 ${[0.261, 0.263]}$ \\ %updated-17-July-2024
 %\hline
 %
 %
 $C_T$ &
 ${4.1~[6.4]}$ & %updated-17-July-2024
 ${-0.033(7)}$ & %updated-17-July-2024
 -- &  $[-0.17,0.17]$ &
 ${[0.350, 0.361]}$ & %updated-17-July-2024
 ${[0.457, 0.473]}$ & %2023 ver check
 ${[0.290, 0.311]}$ \\ %updated-17-July-2024
\bhline{1 pt}
 \end{tabular}
 %}
 \end{adjustbox}
 %\addtolength{\tabcolsep}{-5pt} % set back to normal
 \caption{
 The fit results of the single NP operator scenarios assuming real WCs.
 The WCs are given at the $\mu_b$ scale. 
 The allowed ranges of WC from the $B_c$ lifetime and the current LHC bounds are also shown (``very loose'' represents very weak bounds).
 Fitted WCs and predictions of $P_\tau^D,~P_{\tau}^{D^\ast}$ and $R_{\Jpsi}$
 are in the range of $\Delta\chi^2 = \chi^2 - \chi^2_\text{best} \leq 1$. 
 }
 \label{tab:fit:singlereal}
\end{table}  
%%%%%%%%%%%%%%%%%%%%%%%%%%%%%%%
%%%Table[end]

Once we allow complex values of WCs,
the complex $C_{V_R}$, $C_{S_L}$, and $C_T$ scenarios improve the fits such as 
\begin{align}
 & C_{V_R}  \simeq +0.01 \pm i\, 0.41 & & \text{Pull} = 4.4 \,,& \label{eq:complex_start} \\ %updated-17-July-2024 
 & C_{S_L}  \simeq -0.79 \pm i\, 0.86 & & \text{Pull} = 4.3 \,,& \label{eq:fit_CSL_complex} \\ %updated-17-July-2024 
 &  C_{T} \simeq +0.02 \pm i\, 0.13 & & \text{Pull} = 3.8  \,,\label{eq:complex_end} &  %updated-17-July-2024
\end{align}
while the complex $C_{V_L}$ and $C_{S_R}$ scenarios give the same Pulls compared with those with the real WC scenarios. 
The complex $C_T$ scenario has a similar Pull with the real $C_T$ case due to the penalty of adding one more parameter.
We note that previously the $C_T$ solution had been disfavored by the $F_L^{D^*}$ measurement \cite{Iguro:2018vqb}.
However, the recent combined data revives the complex $C_T$ solution: 
The fitted WC of $C_T$ in Eq.~\eqref{eq:complex_end} predicts $F_L^{D^*}\simeq 0.41$ which is consistent with the measured value within $2\sigma$ level, and hence the near future data will be crucial to test the scenario.
The complex $C_{V_R}$ result at the above best-fit point is, however, not consistent with the LHC bound for the case of EFT, $|C_{V_R}|<0.33$. 
Nevertheless, it could be relaxed in some LQ models with the mass of the LQ particle to be $M_\text{LQ} \gtrsim 2\,\text{TeV}$ as seen in Table~\ref{tab:LHC_bound_one}.

As for the complex $C_{S_L}$ scenario, 
since the imaginary part does not interfere with the SM contribution, the situation is different from the real $C_{S_L}$ scenario, and the $R_{D^{(\ast)}}$ anomalies can be explained.
It, however, looks that the best-fit point in Eq.~(\ref{eq:fit_CSL_complex}) is disfavored by the LHC and $B_c$ lifetime constraints.
Yet, it is noted that the LHC bound is not always proper and depends on the NP model. 
In the case of the charged-Higgs model, for instance, the bound on the $s$-channel mediator $H^\pm$ significantly depends on the resonant mass.
Experimentally, it is not easy to probe the low mass $\tau\nu$ resonance due to the huge SM $W$ background.
Reference~\cite{Iguro:2022uzz} points out that the range of $m_t\leq m_{H^\pm}\leq 400\,\text{GeV}$ is still viable for the $1\sigma$ explanation, 
although LHC Run\,2 data is already enough to probe this range if the $\tau\nu+b$ and $t+\tau\tau$ signatures are searched~\cite{Blanke:2022pjy,Iguro:2023jju}. 
Thus, we leave the LHC bound for the complex $C_{S_L}$ scenario below. 
Once the $B_c$ bound of Eq.~\eqref{eq:Bc} 
 with $\mathcal{B}(B_c \to \tau \overline{\nu})_{\rm UB}=0.6$ is imposed, we find
\begin{align}
 &  C_{S_L}  \simeq -0.57 \pm i\, 0.86 & & \text{Pull} = 4.3\,,& \label{eq:fit_CSL_complex_withBc} %updated-17-July-2024
\end{align}
for the best Pull within the constraint. 
The same pull is obtained as in Eq.~\eqref{eq:fit_CSL_complex} within the digit that we consider.
%%%%%%%%%%%%%%%%%%%%%%%%
%How to obtain (RW): 
%by taking a list of $(Re C_S, Im C_S)$ which is allowed by $B_c$, make a list for $\chi^2(C_S)$ and then find $\chi^2_\text{min}(C_S^\text{best})$.
%%%%%%%%%%%%%%%%%%%%%%%%

It has been pointed out that $q^2$ distributions in $d\Gamma(\Bb \to D^{(\ast)} \tau \overline{\nu})/d q^2$ and $d R_{D^{(\ast)}}/d q^2$ are sensitive to the scalar contribution \cite{Sakaki:2014sea,Celis:2016azn}. 
Furthermore, it is pointed out that a LFU ratio similar to $R_{D^{(*)}}$ with the $q^2$ integration starting from $m_\tau^2$ commonly in both $\tau$ and $\ell$ modes, so-called $\widetilde{R}_{D^{(*)}}$ will be significant \cite{Tanaka:1994ay,Freytsis:2015qca}.
We do not consider the constraint since the experimental data is not conclusive. 
In any case, the Belle II data will be important~\cite{Kou:2018nap}.

%%%Figure[begin]
\begin{figure}[t]
\begin{center}
 \includegraphics[width=1 \textwidth]{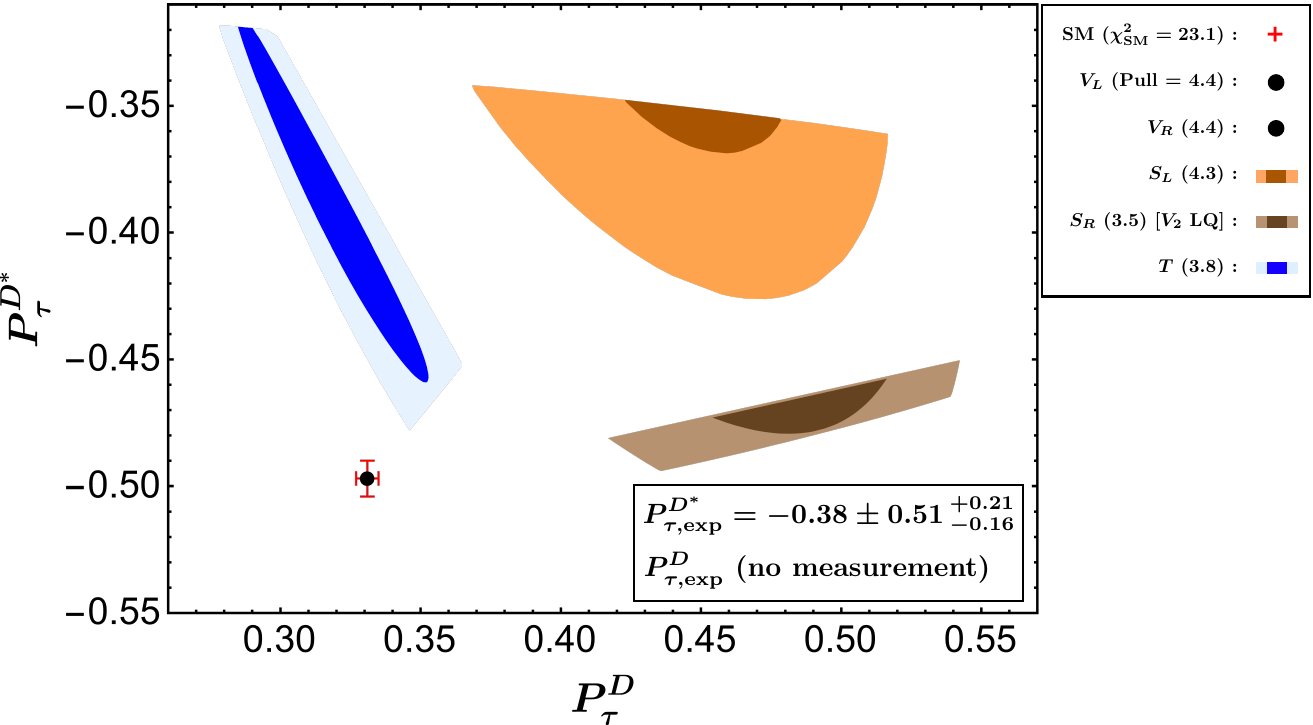}
\end{center}
 \caption{
 \label{fig:EFTprediction}
 Predictions of $P_\tau^D$ and $P_{\tau}^{D^\ast}$ in the complex NP operator scenarios. 
 The allowed regions satisfying $\Delta\chi^2 \le1$ ($4$)  are shown in dark- (light-) orange, brown, and blue for the $C_{S_L}$, $C_{S_R}$, and $C_T$ scenarios, respectively, 
 whereas the black dot is the case for the $C_{V_{L,R}}$ scenarios. 
 The $B_c$ lifetime and LHC bounds are also taken into account. 
The LHC bound is not taken as discussed in the main text. 
}  
\end{figure}
%%%Figure[end]
In Fig.~\ref{fig:EFTprediction}, we show predictions on the plane of $P_\tau^D$--$P_\tau^{D^*}$ evaluated from our fit analysis with each {\it complex} WC scenario.
The allowed regions satisfying $\Delta\chi^2 \le1$ ($4$) are shown in dark- (light-) orange, brown, and blue for the complex $C_{S_L}$, $C_{S_R}$, and $C_{T}$ scenarios, respectively, 
where the $B_c$ lifetime and LHC bounds based on the EFT framework are also taken into account.
The $C_{V_{L,\,R}}$ scenarios do not deviate $P_\tau^D$ and $P_\tau^{D^*}$ from the SM predictions as shown with the black dot in the figure. 
Also note that each shaded region is based on different Pull values, implying  different significance, 
in Fig.~\ref{fig:EFTprediction}. 
We can see that the correlation in $\tau$ polarization observables provides unique predictions that can identify the NP scenarios. 
On the other hand, $R_{\Jpsi}$ is less helpful to distinguish the different operators.

%%%%%%%%%%%%%%%%%%%%%%%%%%%%%%%
\subsection{Leptoquark scenarios} 
%%%%%%%%%%%%%%%%%%%%%%%%%%%%%%%
Finally, we study several LQ scenarios. 
It is well known that three categories of LQs can address the $R_{D^{(*)}}$ anomalies~\cite{Sakaki:2013bfa}, which are referred to as a $SU(2)_L$-singlet vector $\text{U}_1^\mu$, a $SU(2)_L$-singlet scalar $\text{S}_1$, and a $SU(2)_L$-doublet scalar $\text{R}_2$. 
The relevant LQ interactions are given in Appendix~\ref{sec:LQint}.

A key feature with respect to the fit is that these LQ scenarios involve three independent couplings relevant for $b \to c \tau \nu$, 
which are encoded in terms of the two independent (and complex in general) WCs as 
\begin{align}
 \text{U}_1^\mu :~ &  C_{V_L} \,, ~C_{S_R} \,, \\
 \text{S}_1 :~ & C_{V_L} \,, ~C_{S_L} = -4C_T \,, \\
 \text{R}_2 :~ &  C_{V_R} \,, ~C_{S_L} = 4C_T \,,
\end{align}
at the LQ scale $\Lambda_\text{LQ} = M_\text{LQ}$. 

In addition, the $SU(2)_L$-doublet vector LQ $\text{V}_2$ forms $C_{S_R}$~\cite{Sakaki:2013bfa}, equivalent to the single $C_{S_R}$ scenario, 
and hence this LQ has now the viable solution as seen in Sec.~\ref{sec:singleWCfit};
\begin{align}
\text{V}_2^\mu :~ &  C_{S_R} \,,
\end{align}
at the LQ scale.
The relations between the WC and $\text{V}_2$ LQ couplings are also described in Appendix~\ref{sec:LQint}.
In Refs.\,~\cite{Cheung:2022zsb,Iguro:2023prq}, phenomenology of the flavor and collider physics of the $\text{V}_2$ LQ scenario is investigated.
It is found that the coupling product which explains $R_{D^{(*)}}$ also modifies $B_u\to\tau\ov \nu$, $B_s\to \tau\ov \tau$ and $B\to K^{(*)} \tau\ov \tau$ in flavor physics, and the $\tau\ov \tau$ final state search at LHC will be the key probe of the model.

The $C_{V_L}$ phase in $|1+C_{V_L}|^2$ can be absorbed~\cite{Iguro:2018vqb} in the flavor process. 
Thus, the absorption of the $C_{V_L}$ phase is irrelevant for the fit within the flavor observables 
and we take $C_{V_L}$ in $\text{U}_1$ and $\text{S}_1$ LQs as real without loss of generality.\footnote{
Now the real $C_{V_L}$ fit to the $R_{D^{(*)}}$ anomalies gives the minimum $|C_{V_L}|$, and thus is less constrained from the LHC data. 
}
As for $C_{V_R}$ in the $\text{R}_2$ LQ, we assume it as pure imaginary from the fact of Eq.~\eqref{eq:complex_start}.
Therefore, the three LQ scenarios of our interest have three degrees of freedom for the fit and the relevant observables, 
and then it is expected that fit results and their predictions could be different from the previous studies.

These years, UV completions of the LQ scenarios have been studied in the literature; 
Refs.~\cite{DiLuzio:2017vat,Greljo:2018tuh,Cornella:2019hct,DiLuzio:2018zxy,Bordone:2017bld,Bordone:2018nbg,Blanke:2018sro,Balaji:2018zna,Balaji:2019kwe,Fuentes-Martin:2020bnh,Fuentes-Martin:2020hvc,Guadagnoli:2020tlx,Dolan:2020doe,King:2021jeo,Iguro:2021kdw,Iguro:2022ozl} for $\text{U}_1$, 
Refs.~\cite{Heeck:2018ntp,Marzocca:2018wcf,Marzocca:2021azj} for $\text{S}_1$, 
Refs.~\cite{Becirevic:2018afm,Babu:2020hun} for $\text{R}_2$, 
and see also Refs.~\cite{Faber:2018qon,Faber:2018afz}.
In the next subsection, we consider the case if the $\text{U}_1$ LQ is induced by a UV completed theory that gives a specific relation to the LQ couplings, 
and see how it changes the fit result. 
Recent re-evaluations on mass differences of the neutral $B$ mesons $\Delta M_d,\,\Delta M_s$, 
(improved by HQET sum rule and lattice calculations~\cite{DiLuzio:2019jyq}), 
would constrain a UV-completed TeV-scale LQ model~\cite{Calibbi:2017qbu,Marzocca:2018wcf,DiLuzio:2018zxy,Cornella:2019hct,Crivellin:2019dwb,Iguro:2022ozl}.
In particular, the ratio $\Delta M_d/\Delta M_s$ provides a striking constraint on the coupling texture of the LQ interactions. 
Here, we comment that a typical UV completion requires a vector-like lepton (VLL) and it induces additional LQ--VLL box diagrams that also contribute to $\Delta M_{d,s}$ destructively. 
This implies that the constraint from $\Delta M_{d,s}$ depends on the vector-like fermion mass spectrum, and hence we do not consider $\Delta M_{d,s}$ further in our analysis.
(Currently, the mass of the third-generation VLL is constrained to be $\gtrsim 0.5\,\text{TeV}$ by collider searches~\cite{DiLuzio:2018zxy,CMS:2022cpe,CMS:2024bni}, which is a milder bound than conventional searches in $\tau h$ and $\tau Z$ decay channels ($\gtrsim 1\,\text{TeV}$)~\cite{ATLAS:2023sbu,CMS:2024bni}.
This is because in such a model the VLL mainly undergoes LQ-mediated three-body decays.)

The LQ mass has been directly constrained as $M_\text{LQ} \gtrsim 1.5\,\text{TeV}$ from the LQ pair production searches~\cite{Sirunyan:2018vhk,Aaboud:2019bye,Aad:2021rrh}. 
Hence we take $M_\text{LQ} = 2\,\text{TeV}$ for our benchmark scale. 
We recap that the WCs are bounded from the $\tau + \text{missing}$ search and, 
as shown in Table~\ref{tab:LHC_bound_one}, the LQ scenarios receive milder constraints than the EFT operators as long as $M_\text{LQ} \le 10\,\text{TeV}$.

The WCs will be fitted at the $\mu_b$ scale in our analysis, and then they are related to the WCs defined at the $\Lambda_\text{LQ} = M_\text{LQ}$ scale. 
The renormalization-group equations (RGEs) (the first matrix below)~\cite{Jenkins:2013wua,Alonso:2013hga,Gonzalez-Alonso:2017iyc} and the LQ-charge independent QCD one-loop matching (the second one)~\cite{Aebischer:2018acj} gives the following relation
\begin{align}
\begin{pmatrix}
C_{V_L} (\mu_b) \\
C_{V_R} (\mu_b) \\
C_{S_L} (\mu_b) \\
C_{S_R}(\mu_b) \\
C_T (\mu_b)  
\end{pmatrix}
&\simeq
\begin{pmatrix}
1 & 0 &0&0&0\\
0 & 1&0&0&0\\
0 & 0&1.82&0&-0.35\\
0 & 0&0&1.82&0\\
0 & 0&-0.004&0&0.83
\end{pmatrix}
\begin{pmatrix}
1.12 & 0 &0&0&0\\
0 & 1.07&0&0&0\\
0 & 0&1.05&0&0\\
0 & 0&0&1.10&0\\
0 & 0&0&0&1.07
\end{pmatrix}
\begin{pmatrix}
C_{V_L} (\Lambda_\text{LQ} %= 2\,\text{TeV}
) \\
C_{V_R} (\Lambda_\text{LQ} %= 2\,\text{TeV}
)  \\
C_{S_L} (\Lambda_\text{LQ} %= 2\,\text{TeV}
)  \\
C_{S_R} (\Lambda_\text{LQ} %= 2\,\text{TeV}
)  \\
C_T (\Lambda_\text{LQ} %= 2\,\text{TeV}
)   
\end{pmatrix} 
\nonumber
\end{align} 
\begin{align} 
%\\
&\simeq 
\begin{pmatrix}
1.12 & 0 &0&0&0\\
0 & 1.07&0&0&0\\
0 & 0&1.91&0&-0.38\\
0 & 0&0&2.00&0\\
0 & 0&0.&0&0.89
\end{pmatrix}
\begin{pmatrix}
C_{V_L} (\Lambda_\text{LQ} %= 2\,\text{TeV}
) \\
C_{V_R} (\Lambda_\text{LQ} %= 2\,\text{TeV}
)  \\
C_{S_L} (\Lambda_\text{LQ} %= 2\,\text{TeV}
)  \\
C_{S_R} (\Lambda_\text{LQ} %= 2\,\text{TeV}
)  \\
C_T (\Lambda_\text{LQ} %= 2\,\text{TeV}
)   
\end{pmatrix} \,, 
\label{eq:RGE2tev}
\end{align} 
with $\Lambda_\text{LQ} = 2\,\text{TeV}$.
Using these numbers, we obtain $C_{S_L}(\mu_b) = -8.9\, C_T(\mu_b)$ and $C_{S_L}(\mu_b) = 8.4\, C_T(\mu_b)$ for $\text{S}_1$ and $\text{R}_2$ LQs, respectively.

With these ingredients, the LQ scenarios in terms of $C_X (\mu_b)$ up to three degrees of freedom are investigated, 
where the full variable case is referred to as the general LQ. 
The results of the best-fit points for the general LQ scenarios that are also allowed from the  $B_c$ and LHC bounds are then summarized as 
\begin{align}
 \label{eq:LQfit_start}
 &\text{U}_1\text{~LQ}:~& 
 & C_{V_L}=0.07\,,~C_{S_R}= 0.02\,,& %updated-17-July-2024
 &\text{Pull} = 4.1\,,&  %updated-17-July-2024
 \\[0.5em]
 &\text{S}_1\text{~LQ}:~& 
 &C_{V_L}= 0.07\,,~C_{S_L}= -8.9\,C_T = \pm i 0.15 \,, & %updated-17-July-2024 
 &\text{Pull} = 4.1\,,&  %updated-17-July-2024
 \\[0.5em]
 \label{eq:R2fit}
 &\text{R}_2\text{~LQ}:~& 
 &C_{V_R}=\pm i 0.50\,,~C_{S_L}= 8.4\,C_T = 0.03 \mp i 0.18\,,& %updated-19-Aug-2024
 &\text{Pull} = 4.1\,.& %updated-17-July-2024
\end{align}
We observe that these three general LQ scenarios have the same Pull, which means they are equivalently favored by the current data.
We see that at the best-fit points the $\text{U}_1$, $\text{S}_1$, and $\text{R}_2$ LQ scenarios prefer purely real, purely imaginary, and complex scalar NP contributions, respectively.

Regarding the general $\text{S}_1$ LQ scenario, we comment that a part of the allowed parameter region is ruled out by the $B \to K^\ast \nu\overline\nu$ measurement and $\Delta M_s$ (via LQ--$\nu_\tau$ box)~\cite{Crivellin:2019dwb,Crivellin:2021lix,Endo:2021lhi}, although these constraints can be avoided by tuning the LQ couplings ($y_L^{b\tau} \gg y_L^{s\tau}$ defined in Eq.~\eqref{eq:S1L} which leads to $C_{V_L}\simeq0$).

Furthermore, we also investigate two restricted LQ scenarios; $\text{S}_1$ LQ with $C_{V_L}=0$ and 
$\text{R}_2$ LQ with $C_{V_R}=0$. The former scenario naturally avoids the severe constraint from $\Delta M_{s}$ without introducing the VLLs \cite{Endo:2021lhi}. 
The latter scenario is also natural in light of the naive dimensional analysis, where the $C_{V_R}$ contribution corresponds to the dimension-eight operator in the EFT as mentioned in Sec.~\ref{sec_GF}.
The fit results for the $\text{S}_1$ LQ with $C_{V_L} =0 $  and the $\text{R}_2$ LQ with $C_{V_R} = 0$ are obtained as 
\begin{align}
 &\text{S}_1\text{~LQ ($C_{V_L} = 0$)}:~& 
 &C_{S_L}= -8.9\,C_T = 0.18\,,& %updated-17-July-2024
 &\text{Pull} = 4.1\,,&  %updated-17-July-2024
 \\[0.5em]
 &\text{R}_2\text{~LQ ($C_{V_R} = 0$)}:~& 
 &C_{S_L}= 8.4\,C_T = -0.09 \pm i 0.56\,,& %updated-17-July-2024
 &\text{Pull} = 4.4\,, %updated-17-July-2024
 \label{eq:LQfit_end}
\end{align}
where the improvements of Pull only come from the benefit of reducing the degrees of freedom.

In turn, we evaluate the LHC bounds
on the two independent variables, such as $(C_{V_L},C_{S_R})$ for the $\text{U}_1$ LQ scenario, by the following interpretations: 
\begin{align}
 &\text{U}_1\text{~LQ}:~~
 \frac{|C_{V_L}(\mu_b)|^2}{(0.42)^2} + \frac{|C_{S_R}(\mu_b)|^2}{(0.77)^2} < 1 \,, & \label{eq:VLSR} \\
 &\text{S}_1\text{~LQ}:~~
 \frac{|C_{V_L}(\mu_b)|^2}{(0.42)^2} + \frac{|C_{S_L}(\mu_b)|^2}{(0.80)^2} < 1 \,, & \label{eq:VLSL} \\
 &\text{R}_2\text{~LQ}:~~
 \frac{|C_{V_R}(\mu_b)|^2}{(0.51)^2} + \frac{|C_{S_L}(\mu_b)|^2}{(0.80)^2} < 1 \,, & \label{eq:VRSL}
\end{align}
where the denominators are the current LHC bounds for the single WC scenarios with $M_\text{LQ} = 2\,\text{TeV}$ from Table~\ref{tab:LHC_bound_one}. 
Indeed, this is a good approximation since the bound comes from the high-$p_T$ region that suppresses the interference term between the $V_{L,R}$ and $S_{L,R}$ operators. 
It can be seen that the best-fit point of Eq.~\eqref{eq:R2fit} for $\text{R}_2$ LQ is not consistent with the LHC bound in  Eq.~\eqref{eq:VRSL}.

%%%Table[begin]
\begin{table}[t]
\centering
\newcommand{\bhline}[1]{\noalign{\hrule height #1}}
\renewcommand{\arraystretch}{1.5}
%\rowcolors{3}{white}{gray!15}
%\addtolength{\tabcolsep}{5pt} % add space between columns
  % \scalebox{0.8}{
   \begin{adjustbox}{width=\columnwidth,center}
  \begin{tabular}{l c c|c c|c c c} 
  \bhline{1 pt}
 \multirow{2}{*}{ } & 
 \multirow{2}{*}{Pull~[$\chi^2_\text{best}$]} & 
 \multirow{2}{*}{Fitted $C_X$} &
 \multicolumn{2}{c|}{Allowed region of $C_X$}&
 \multicolumn{3}{c}{Predictions $(\Delta\chi^2 \le 1)$} \\
 %\hline
 %
 & & %$(\Delta \chi^2 \le 1)$ 
 & $B_c \to \tau \overline \nu$ & LHC & $P_\tau^D$ & $-P_\tau^{D^*}$ & $R_{\Jpsi}$  \\
 \hline
 SM &
 --${~[23.1]}$ &
 -- &
 -- &  -- &
 $0.331 \pm 0.004$ &
 $0.497 \pm 0.007$ &
 $0.258 \pm 0.004$ \\
 \hline
 %
 %
 %%%%%%%%%%%%%%%%% U1LQ
 \multirow{3}{*}{$\text{U}_1$ LQ} & 
 \multirow{3}{*}{$4.1~[0.2]$} & %updated-17-July-2024
 $~~~\,C_{V_L}:\,[+0.048,+0.104]$ & %updated-17-July-2024
 \multirow{3}{*}{Eq.~\eqref{eq:Bc}} & \multirow{3}{*}{Eq.~\eqref{eq:VLSR}} &
 \multirow{3}{*}{$[0.272, 0.413]$} & %updated-17-July-2024
 \multirow{3}{*}{$[0.482, 0.533]$} & %updated-17-July-2024
 \multirow{3}{*}{$[0.285, 0.311]$} \\ %updated-17-July-2024
 %\hline
 & 
 &
 $\text{Re}C_{S_R}:\,[-0.058,+0.090]$ & %updated-17-July-2024
 &
 &
 \\
 & 
 &
 $\text{Im}C_{S_R}:\, [-0.390,+0.390]$ & %updated-17-July-2024
  &
 &
 \\ \hline
 %%%%%%%%%%%%%%%%% U1LQ
 %
%%%%%%%%%%%%%%%%% S1LQ
 \multirow{3}{*}{$\text{S}_1$ LQ} & 
 \multirow{3}{*}{$4.1~[0.3]$} & %updated-17-July-2024
 $~~~\,C_{V_L}:\,[+0.032,+0.160]$ & %updated-17-July-2024
 \multirow{3}{*}{Eq.~\eqref{eq:Bc}} & \multirow{3}{*}{Eq.~\eqref{eq:VLSL}} &
 \multirow{3}{*}{$[0.086, 0.442]$} & %updated-17-July-2024
 \multirow{3}{*}{$[0.374, 0.502]$} & %updated-17-July-2024
 \multirow{3}{*}{$[0.275, 0.312]$} \\ %updated-17-July-2024
 %\hline
 & 
 &
 $\text{Re}C_{S_L}:\,[-0.110,+0.110]$ & %updated-17-July-2024
 &
 &
 \\
 & 
 &
 $\text{Im}C_{S_L}:\, [-0.416,+0.416]$ & %updated-17-July-2024
  &
 &
 \\ \hline
 %%%%%%%%%%%%%%%%% S1LQ
  %
 %%%%%%%%%%%%%%%%% S1LQ CV=0
 %\multirow{2}{*}{$\text{R}_2$ LQ} & 
 $\text{S}_1$ LQ &
 \multirow{2}{*}{$4.1~[3.0]$} & %updated-17-July-2024
 $\text{Re}C_{S_L}:\,[+0.014,+0.210]$ & %updated-17-July-2024
 \multirow{2}{*}{Eq.~\eqref{eq:Bc}} &  
 \multirow{2}{*}{Eq.~\eqref{eq:VRSL}} &
 \multirow{2}{*}{$[0.466, 0.524]$} & %updated-17-July-2024
 \multirow{2}{*}{$[0.456, 0.504]$} & %updated-17-July-2024
 \multirow{2}{*}{$[0.271, 0.284]$} \\ %updated-17-July-2024
 %\hline
 $(C_{V_L}=0)$ & 
 &
 $\text{Im}C_{S_L}:\,[-0.522, +0.522]$ & %updated-17-July-2024
 &  &
 &
 \\ \hline
 %%%%%%%%%%%%%%%%% S1LQ CV=0
 %
 %%%%%%%%%%%%%%%%% R2LQ
 \multirow{3}{*}{$\text{R}_2$ LQ} & 
 \multirow{3}{*}{$4.1~[0.1]$} & %updated-17-July-2024
 $\text{Im}C_{V_R}:\,[\pm 0.000,\pm 0.504]$ & %updated-17-July-2024
 \multirow{3}{*}{Eq.~\eqref{eq:Bc}} & \multirow{3}{*}{Eq.~\eqref{eq:VRSL}} &
 \multirow{3}{*}{$[0.259, 0.478]$} & %updated-17-July-2024
 \multirow{3}{*}{$[0.402, 0.533]$} & %updated-17-July-2024
 \multirow{3}{*}{$[0.280, 0.313]$} \\ %updated-17-July-2024
 %\hline
 & 
 &
 $\text{Re}C_{S_L}:\,[-0.036,+0.054]$ & %updated-17-July-2024
 &
 &
 \\
 & 
 &
 $\text{Im}C_{S_L}:\, [\mp 0.000, \mp 0.310]$ & %updated-17-July-2024
  &
 &
 \\ \hline
 %%%%%%%%%%%%%%%%% R2LQ
 %
 %%%%%%%%%%%%%%%%% R2LQ CV=0
 %\multirow{2}{*}{$\text{R}_2$ LQ} & 
 $\text{R}_2$ LQ &
 \multirow{2}{*}{$4.4~[0.6]$} & %updated-17-July-2024
 $\text{Re}C_{S_L}:\,[-0.148,-0.042]$ & %updated-17-July-2024
 \multirow{2}{*}{Eq.~\eqref{eq:Bc}} &  
 \multirow{2}{*}{Eq.~\eqref{eq:VRSL}} &
 \multirow{2}{*}{$[0.404, 0.479]$} & %updated-17-July-2024
 \multirow{2}{*}{$[0.408, 0.440]$} & %updated-17-July-2024
 \multirow{2}{*}{$[0.278, 0.299]$} \\ %updated-17-July-2024
 %\hline
 $(C_{V_R}=0)$ & 
 &
 $\text{Im}C_{S_L}:\,[\pm 0.503,\pm 0.619]$ & %updated-17-July-2024
 &  &
 &
 \\ \hline
 %%%%%%%%%%%%%%%%% R2LQ CV=0
 %
\bhline{1 pt}
 \end{tabular}
 %}
 \end{adjustbox}
 \caption{
 The fit results of the $\text{U}_1$, $\text{S}_1$, and $\text{R}_2$ LQ scenarios for $M_\text{LQ} = 2\,\text{TeV}$. 
 The WCs are given at the $\mu_b$ scale, whose allowed ranges are cut by the $B_c$ lifetime and the LHC bounds.
 The structure is the same as in Table~\ref{tab:fit:singlereal}.
 }
 \label{tab:fit:leptoquark}
\end{table}  
%%%%%%%%%%%%%%%%%%%%%%%%%%%%%%%
%%%Table[end]

In Table~\ref{tab:fit:leptoquark}, we show our fit results and predictions with respect to each LQ scenario as we did for the EFT cases. 
It is observed that both the general LQ scenarios and the restricted LQ scenarios can largely deviate the $\tau$ polarizations from the SM predictions. 
This can be understood from the fact that the complex scalar WCs have large impacts on the interference terms, as can be checked from Eqs.~\eqref{eq:PtauD} and \eqref{eq:PtauDs}, 
which results in a wide range of predictions.

\begin{figure}[t]
\begin{center}
 \includegraphics[width=1 \textwidth]{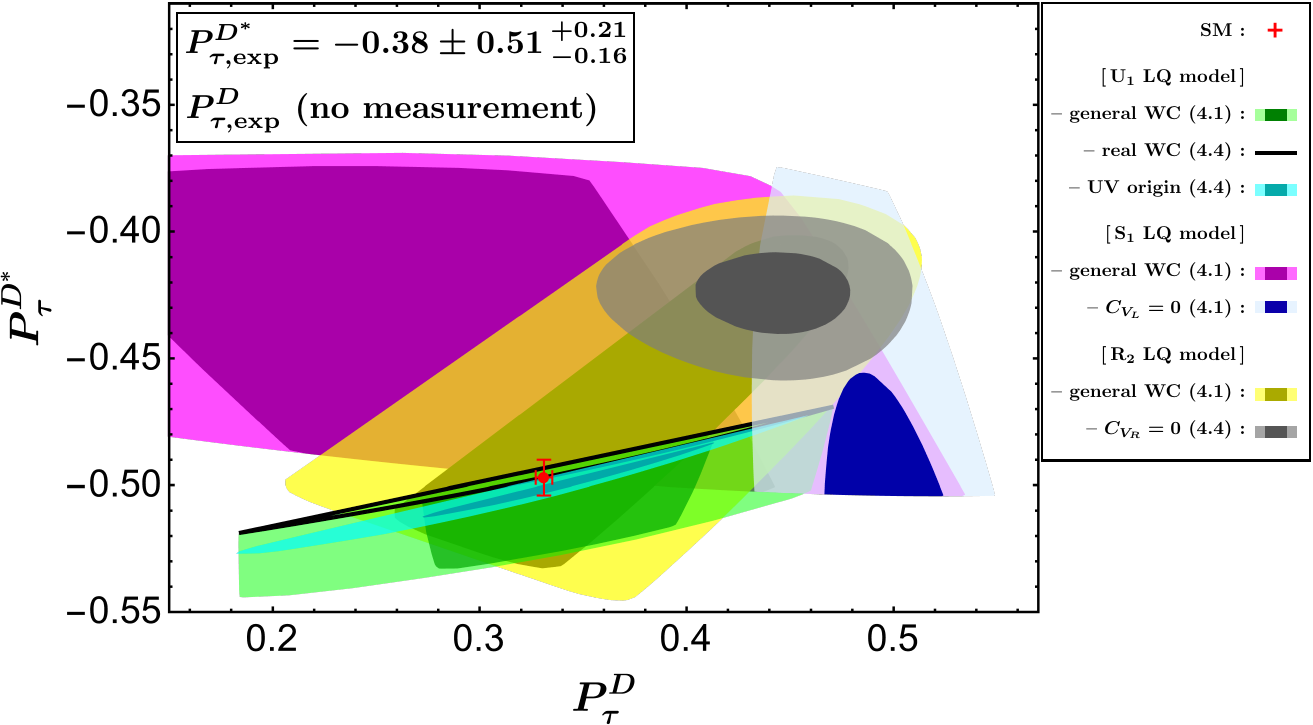} %updated-17-July-2024
\end{center}
 \caption{
 Predictions of $P_\tau^D$ and $P_{\tau}^{D^\ast}$ in the LQ scenarios following the same procedure as in Fig.~\ref{fig:EFTprediction}. 
 The allowed regions are shown by dark (light) green, magenta, and yellow regions for the general $\text{U}_1$, $\text{S}_1$, and $\text{R}_2$ LQ scenarios, respectively. 
 The specific scenarios; 
 $\text{U}_1$ LQ with UV origin (cyan), $\text{U}_1$ LQ with real WCs (solid line),
 $\text{S}_1$ LQ with $C_{V_L}=0$ (blue), and
 $\text{R}_2$ LQ with $C_{V_R}=0$ (gray), are also shown.  
 \label{fig:LQprediction}
 } 
\end{figure}

In Fig.~\ref{fig:LQprediction}, we show the combined the $\tau$ polarization predictions on the $P_\tau^D$--$P_\tau^{D^*}$ plane satisfying $\Delta\chi^2 \le1\,(4)$ and the aforementioned $ B_c$ lifetime and the LHC bounds, where the general $\text{U}_1$, $\text{S}_1$, and $\text{R}_2$ LQ scenarios are shown in dark (light) green, magenta, and yellow regions, respectively. 
The $\text{U}_1$ and $\text{R}_2$ LQ scenarios produce the correlated regions of the $P_\tau^D$--$P_\tau^{D^*}$ predictions and hence could be distinguished. 
On the other hand, the $\text{S}_1$ LQ scenario has a less-predictive wide region, which is hard to be identified.

Figure~\ref{fig:LQprediction} also exhibits the predictions for several specific LQ scenarios, \ie, 
$\text{U}_1$ LQ with real $C_{S_R}$ \& $C_{V_L}$ (solid line), $\text{S}_1$ LQ with $C_{V_L}=0$ (blue region), and $\text{R}_2$ LQ with $C_{V_R}=0$ (gray region).  
It is seen that reducing the variable in the general LQ scenario provides the distinct prediction in particular for $P_\tau^{D^*}$ and 
the correlation for $P_\tau^D$--$P_\tau^{D^*}$ becomes a useful tool to identify the LQ signature.
Therefore, it is significant to restrict the LQ interactions by the $\tau$ polarization observables
or by constructing a UV theory that realizes the LQ particle. 
The latter will be discussed in the next section for the $\text{U}_1$ LQ scenario, which corresponds to the cyan region in Fig.~\ref{fig:LQprediction}.

%%%Figure[begin]
%%%Figure[end]

%%%%%%%%%%%%%%%%%%%%%%%%%%%%%%%
\subsection{UV completion of \texorpdfstring{$\text{U}_1$}{U1} leptoquark}
%%%%%%%%%%%%%%%%%%%%%%%%%%%%%%%
As the $\text{U}_1$ LQ provides a unique solution, not only to the $b \to c \tau \nu$ anomaly but also to several flavor issues, UV completions of the $\text{U}_1$ LQ have been discussed enthusiastically~\cite{Barbieri:1995uv,Barbieri:1997tu,Barbieri:2011ci,Barbieri:2011fc,Barbieri:2012uh,Blankenburg:2012nx,Barbieri:2015yvd,Fuentes-Martin:2019mun,FernandezNavarro:2022gst}.  
A typical description is that the $\text{U}_1$ LQ is given as a new gauge boson, embedded in a large gauge symmetry, 
such that the third-generation quarks and leptons are coupled to $\text{U}_1$ in the interaction basis. 
This means that the two LQ interactions of Eq.~\eqref{eq:U1int} are represented as a universal gauge coupling, $x_L^{33} = x_R^{33} \equiv g_U$ (see Appendix~\ref{sec:LQint}). 
Moving to the mass basis leads to 
\begin{align}
 \label{eq:U2WC_NPscale}
 C_{S_R} (\Lambda_{\text{LQ}})= -2 \beta_R \times C_{V_L} (\Lambda_{\text{LQ}}) \,,     
\end{align}
where $\beta_R = e^{i \phi}$ denotes the relative complex (CP-violating) phase~\cite{Fuentes-Martin:2019mun}, 
which comes from the fact that the phases in the rotation matrices (to the mass basis) for quark and lepton are not necessarily identical. 
The LHC bound for this scenario has been studied and the typical scale of the constraint is obtained as $\Lambda_{\text{LQ}} \gtrsim 3.5\,\text{TeV}$~\cite{Cornella:2019hct}.

The RGE running effect changes the above relation of Eq.~\eqref{eq:U2WC_NPscale} at the $\mu_b$ scale of our interest. 
By taking $\Lambda_{\text{LQ}} = 4\,\text{TeV}$ as a benchmark scale, we obtain 
\begin{align}
\begin{aligned}
C_{V_L} (\mu_b) &= 1\times 1.11 \times C_{V_L} (\Lambda_{\text{LQ}} %= 4\,\text{TeV}
) \,,\\
C_{S_R}(\mu_b) &= 1.90 \times 1.09 \times C_{S_R} (\Lambda_{\text{LQ}}% = 4\,\text{TeV}
) \,,
\label{eq:RGE4tev}
\end{aligned}    
\end{align}
where the first coefficient is the QCD two-loop RGE factor~\cite{Gonzalez-Alonso:2017iyc} and the second is the QCD one-loop matching correction~\cite{Aebischer:2018acj} at the NP scale.
Therefore, we have 
\begin{align}
 C_{S_R}(\mu_b) \simeq -3.7\, \beta_R \times C_{V_L}(\mu_b) \,, 
\end{align}
in the case of the UV origin $\text{U}_1$ LQ scenario, applied to our fit analysis.

The result of the best-fit point for the UV origin $\text{U}_1$ LQ scenario, with the definition of $\beta_R = e^{i \phi}$, is shown as 
\begin{align}
\label{eq:U2result}
& (C_{V_L}, \phi) \,\simeq\, (0.075\,, \pm 0.466\pi) & & \text{Pull} = 4.4\,.& %updated-17-July-2024
\end{align}
One can see that this is consistent with the $B_c$ lifetime and LHC bounds. 
Predictions of the observables within $\Delta\chi^2 \le 1,4$ are shown by the cyan region in Fig.~\ref{fig:LQprediction}. 
It is observed that the large complex phase is favored, which suppresses the interference while predicting nucleon electric dipole moments ($d_N$) within the reach of future experiments~\cite{Iguro:2023rom}.
It should also be stressed that the $\tau$ polarizations are so unique that this scenario can be distinguished from the aforementioned LQ scenarios.

As briefly mentioned in the previous section, although the simplified model is severely constrained from the $\Delta M_{s}$ measurement (when ${x}^{s\tau}_L \neq 0$),
it can be avoided by the existence of the VLL.  There is a GIM-like mechanism including the VLL contributions, so that the bound from $\Delta M_{s}$ is naturally suppressed even when the VLL mass is around 1--4 TeV \cite{Calibbi:2017qbu,Cornella:2019hct,DiLuzio:2018zxy,Fuentes-Martin:2020hvc,Marzocca:2018wcf,Crivellin:2018yvo,Iguro:2022ozl}.

%%%%%%%%%%%%%%%%%%%%%%%%%%%%%
\section{The LFU violation in \texorpdfstring{\boldmath{$\Upsilon$}}{Upsilon} decays}
\label{sec:Upsi_decay}
%%%%%%%%%%%%%%%%%%%%%%%%%%%%%

The UV-completed NP models 
contributing to  $b \to c \tau \overline\nu$ processes should also bring a related contribution to
$b \overline{b}\to \tau^+ \tau^-$ or $c \overline{c} \to \tau^+ \tau^-$ interactions \cite{Aloni:2017eny, Garcia-Duque:2021qmg,Garcia-Duque:2022tti}.
In this section, we show that
 $\text{U}_1$ and $\text{R}_2$ LQs predict a robust correlation 
between $b \to c \tau \overline\nu$ and $b \overline{b}\to \tau^+ \tau^-$ via the LQ exchange.

A definition of the LFU observable in the $\Upsilon (nS)$ decays is  
\beq
R_{\Upsilon(nS)}\equiv  \frac{\mathcal{B}(\Upsilon(nS) \to \tau^+ \tau^-)}{\mathcal{B}(\Upsilon(nS) \to \ell^+ \ell^-)}\,,
\eeq
with $n=1,2,3$, where $R_{\Upsilon (nS)} \simeq 1$ holds in the SM. 
As for $n\geq 4$, the leptonic branching ratios are significantly suppressed since a $B\Bb$ decay channel is open.%
\footnote{
A novel method for the $n=4$ mode has been proposed in Ref.~\cite{Genon} by using the inclusive di-leptonic channel 
$\Upsilon(4S) \to \ell^\pm \tau^\mp X (\overline\nu \nu)$, 
which could be probed in the Belle II experiment and is directly related to
$\Gamma (b \to X \tau \nu)/\Gamma(b \to X \ell \nu)$.
}
Since the short- and long-distance QCD corrections \cite{Beneke:2014qea} are independent of the lepton mass, they are canceled in this ratio. 
One can also discuss the $c \overline{c} \to l^+ l^-$ LFU observable via $\psi(2S)$ decays.
However, we do not consider it because the present experimental error is relatively large.

Recently, the BaBar collaboration has reported a precise result for the measurement of  $R_{\Upsilon(3S)}$ \cite{Lees:2020kom}:  
$
R_{\Upsilon(3S)}^{\rm BaBar} = 0.966 \pm 0.008_{\rm stat} \pm 0.014_{\rm syst},
$
where $\ell = \mu$.
Combing a previous measurement by the  CLEO  collaboration  \cite{Besson:2006gj}, 
an average for the $\Upsilon (3S)$ decay is  
\cite{Garcia-Duque:2021qmg}
\beq
R_{\Upsilon(3S)}^{\rm exp} = 0.968 \pm 0.016\,.
\label{eq:Upsilonexp}
\eeq
This value is consistent with the SM prediction \cite{Aloni:2017eny}
\beq
R_{\Upsilon(3S)}^{\rm SM}  = 0.9948 \pm \mathcal{O}(10^{-5})\,,
\eeq 
at the $1.7\sigma$ level.
The SM prediction slightly deviates from $1$ whose leading correction comes from the difference in the phase space factor between the $\tau/\ell$ modes \cite{VanRoyen:1967nq}.
The next-to-leading contribution comes from the QED correction, which depends on the lepton mass~\cite{Bardin:1999ak};  
$\delta_{\rm EM} R_{\Upsilon(nS)}=+0.0002$. 
The tree-level $Z$ exchange also contributes, but its effect is $\mathcal{O}(10^{-5})$ \cite{Aloni:2017eny}.
There is no Higgs boson contribution, as one can see below.
The other channels ($n=1,2$) still suffer from
the current experimental uncertainty,
and we do not utilize them in our presentation.

The effective Hamiltonian which is relevant to the bottomonium decay into $\tau^+ \tau^-$ is described as
\beq
- \mathcal{H}^{\rm NP}_{\rm eff} = & ~ C_{VLL}^{b \tau} (\overline{b}\gamma^{\mu}P_L b) (\overline{\tau}\gamma_{\mu} P_L \tau)
+ C_{VRR}^{b \tau} (\overline{b}\gamma^{\mu}P_R b) 
(\overline{\tau}\gamma_{\mu} P_R \tau) \nonumber \\
 & + C_{VLR}^{b \tau} (\overline{b}\gamma^{\mu}P_L b) (\overline{\tau}\gamma_{\mu} P_R \tau)
+C_{VRL}^{b \tau} (\overline{b}\gamma^{\mu}P_R b) (\overline{\tau}\gamma_{\mu} P_L \tau) \\
& + \left[ C_T^{b \tau} (\overline{b} \sigma^{\mu \nu} P_R b )(\overline{\tau} \sigma_{\mu \nu} P_R \tau ) 
+ C_{SL}^{b \tau}  (\overline{b}P_L b)(\overline{\tau}P_L \tau)
+ C_{SR}^{b \tau}  (\overline{b}P_R b)(\overline{\tau}P_L \tau)
+ \textrm{h.c.}\right]\,,
\nonumber 
\eeq
at the scale $\mu = m_{\Upsilon}$. Note that $C_{VLL}^{b \tau}, C_{VRR}^{b \tau}, C_{VLR}^{b \tau}$ and $C_{VRL}^{b \tau}$ are real coefficients, and  
$C_{SL}^{b \tau}$ and $C_{SR}^{b \tau} $ never contribute to the $\Upsilon(nS) \to \tau^+\tau^-$ due to $\langle 0 | \overline{b} b | \Upsilon \rangle =\langle 0 | \overline{b}\gamma_5 b | \Upsilon \rangle=0$.
In this convention, the partial decay width is given by \cite{Aloni:2017eny}
\beq
\begin{aligned}
\Gamma(\Upsilon(nS) \to \tau^+ \tau^-)= &~
\frac{f_\Upsilon^2}{4 \pi m_\Upsilon}
\sqrt{1 - 4 x_{\tau}^2} \Bigl[ 
 A_\Upsilon^2 (1 + 2 x_{\tau}^2) 
+ B_\Upsilon^2 (1 - 4 x_{\tau}^2) 
  \\
& \quad + \frac{1}{2}C_\Upsilon^2  
(1 - 4 x_{\tau}^2)^2 + \frac{1}{2} D_\Upsilon ^2  
(1 - 4 x_{\tau}^2)  + 2   A_\Upsilon 
C_\Upsilon x_{\tau} (1 - 4 x_{\tau}^2)\Bigr]\,,
\end{aligned}
\eeq
with
\beq
A_\Upsilon =& \frac{4 \pi \alpha}{3} + 
\frac{m_{\Upsilon}^2}{4} \left[  C_{VLL}^{b \tau}+  C_{VRR}^{b \tau}+ C_{VLR}^{b \tau}+C_{VRL}^{b \tau}+ 16  x_{\tau} \frac{f_\Upsilon^T}{f_\Upsilon} \textrm{Re}\left(C_T^{b\tau}\right)\right]\,,\\
B_\Upsilon =& \frac{m_{\Upsilon}^2}{4} 
 \left(  C_{VRR}^{b \tau}+ C_{VLR}^{b \tau} -C_{VLL}^{b \tau} -C_{VRL}^{b \tau}\right)\,,\\
C_\Upsilon = & 2 m_\Upsilon^2 \frac{f_\Upsilon^T}{f_\Upsilon}
\textrm{Re} \left( C_T^{b \tau} \right)\,,\\
D_\Upsilon =& 2 m_\Upsilon^2 \frac{f_\Upsilon^T}{f_\Upsilon} 
\textrm{Im} \left( C_T^{b\tau} \right)\,,
\eeq
and 
\beq
x_\tau =\frac{m_{\tau}}{m_{\Upsilon}}\,.
\eeq
The $f_{\Upsilon}$ and $f_{\Upsilon}^T$ are decay constants for vector and tensor currents in $\Upsilon$ hadronic-matrix elements, 
and $f_{\Upsilon} = f_{\Upsilon}^T$ holds in the heavy quark limit, which is realized for the $\Upsilon$ decays \cite{Aloni:2017eny}.

Within the SM, this process is predominantly caused by the QED.
Nevertheless, the photon-exchange QED contribution is suppressed by $1/m_{\Upsilon}^2 $, 
and hence the NP contribution could be non-negligible~\cite{Aloni:2017eny,Matsuzaki:2018jui,Garcia-Duque:2021qmg}.
In the SM, 
$A_\Upsilon \simeq 4 \pi \alpha/3$ and $B_\Upsilon, C_\Upsilon, D_\Upsilon \simeq 0$. 
Setting the light lepton mass to zero and $m_\Upsilon = m_{\Upsilon(3S)} = 10.355\,\text{GeV}$, 
we obtain the following numerical formula 
\beq
\begin{aligned}
\frac{R_{\Upsilon(3S)}}{R_{\Upsilon(3S)}^{\rm SM}} =&~  
1 +1.64\times 10^{-3}\ {\rm TeV}^2 \left( C_{VLL}^{b \tau}+  C_{VRR}^{b \tau}+ C_{VLR}^{b \tau}+C_{VRL}^{b \tau}\right) \\
& + 6.37\times 10^{-3}\ {\rm TeV}^2\  \textrm{Re}\left(C_T^{b \tau} \right)+ \delta_\Upsilon\,, 
\end{aligned}
\eeq
with
\beq
\begin{aligned}
\delta_\Upsilon = &~ 5.22 \times 10^{-6}\ {\rm TeV}^4 \left( C_{VLL}^{b \tau}+  C_{VRR}^{b \tau}+ C_{VLR}^{b \tau}+C_{VRL}^{b \tau}\right)\textrm{Re}\left(C_T^{b \tau} \right)
 \\
&+ 6.71 \times 10^{-7}\ {\rm TeV}^4 \left( C_{VLL}^{b \tau}+  C_{VRR}^{b \tau}+ C_{VLR}^{b \tau}+C_{VRL}^{b \tau}\right)^2  
\\
&+5.59 \times 10^{-7}\ {\rm TeV}^4 \left(  C_{VRR}^{b \tau}+ C_{VLR}^{b \tau} -C_{VLL}^{b \tau} -C_{VRL}^{b \tau}\right)^2 
 \\
& + 2.51 \times 10^{-5}\ {\rm TeV}^4 \left[\textrm{Re}\left(C_T^{b \tau} \right)\right]^2  
+  1.79 \times 10^{-5}\ {\rm TeV}^4 \left[\textrm{Im}\left(C_T^{b \tau} \right)\right]^2\,,
\end{aligned}
\eeq
where the $\delta_\Upsilon$ term gives negligible contributions.

Let us now look into a correlation between 
$R_{\Upsilon(3S)}$ and $R_{D^{(\ast)}}$ by using the specific examples of the $\text{U}_1$ and $\text{R}_2$ LQs.
First, we exhibit the $\text{U}_1$ LQ case.
The $\text{U}_1$ LQ interaction with the SM fermions is given in Eq.~\eqref{eq:U1int}.
Integrating the $\text{U}_1$ LQ out, 
as well as the charged current contributions ($b \to c \tau \overline\nu$) in Eq.~\eqref{eq:U1charged}, 
the neutral current ones ($b \overline{b} \to \tau^+ \tau^-$) are  obtained as 
\beq
C_{VLL}^{b\tau}(\mu_{\rm LQ}) %=C_{VLL}^{b\tau}(m_{\Upsilon}) 
= -\frac{|x_L^{b\tau}|^2}{m_{U_1}^2}\,,
\quad
C_{VRR}^{b\tau}(\mu_{\rm LQ}) %=C_{VRR}^{b\tau}(m_{\Upsilon})
= -\frac{|x_R^{b\tau}|^2}{m_{U_1}^2}\,,
\quad
C_{SR}^{b \tau} (\mu_{\rm LQ}) 
= \frac{2 x_L^{b \tau} (x_R^{b \tau})^\ast}{m_{U_1}^2}\,.
\eeq
The vector contributions do not change under the RGEs, while the scalar contribution does not affect the $\Upsilon$ decay.
Here,
an important point is that $R_{\Upsilon(3S)} / R_{\Upsilon(3S)}^{\rm SM}$ 
is predicted to be less than $1$ when NP contributions are dominated by vector interactions.
It would lead to a coherent deviation with $R_{D^{(\ast)}}$.
%%%%%%%%%%%%%%%%%%%%%%
\begin{figure}[t]
    \centering
    \includegraphics[width=0.7\textwidth]{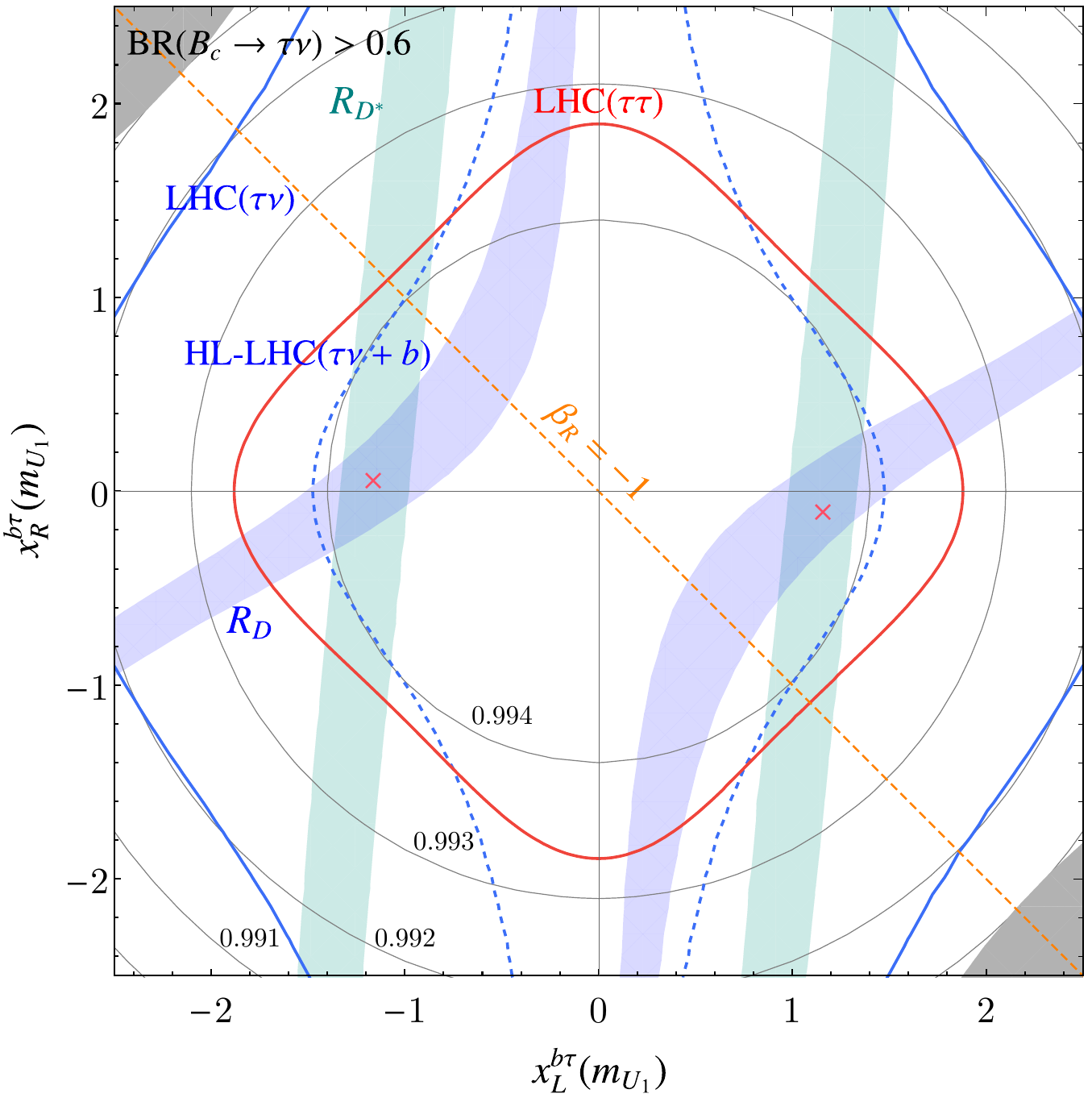}
    \caption{
    The correlation between $R_{\Upsilon(3S)}$ and $R_{D^{(\ast)}}$ is exhibited in the $\text{U}_1$ LQ scenario with 
    $m_{\text{U}_1}=2\,$TeV by setting $x_L^{s \tau}/x_L^{b \tau} =\lambda$. 
    The predicted values of $R_{\Upsilon(3S)}$ are shown by black contours.
    $R_D$ and $R_{D^\ast}$ anomalies can be explained in the blue and green regions, respectively.
    The LHC exclusion region from the $\tau+\,$missing searches (outside the blue line) and the HL-LHC sensitivity (blue dashed line) are based on the result of Table~\ref{tab:LHC_bound_one}.
    The LHC exclusion region from the non-resonant $\tau^+\tau^-$ searches is shown by the red line.
    The best-fit points of Eq.~\eqref{eq:LQfit_start} are shown by red crosses. The orange dashed line represents the $U(2)$ flavor symmetry prediction with $\beta_R = -1$. 
    The gray-shaded region is excluded by the $B_c$ lifetime. 
     \label{fig:U1Upsilon}
    }
\end{figure}
%%%%%%%%%%%%%%%%%%%%%%

Setting $m_{U_1}=2.0\,\text{TeV}$ and $(V x_L (\mu_{\text{LQ}}))^{c \tau}  = V_{cb} x_L^{b \tau} (\mu_{\text{LQ}}) + V_{cs} x_L^{s \tau} (\mu_{\text{LQ}}) $ with $x_L^{s \tau}/x_L^{b \tau} =\lambda \simeq 0.225$,  
we show a correlation between $R_{\Upsilon(3S)}$ and $R_{D^{(\ast)}}$ in Fig.~\ref{fig:U1Upsilon} assuming $x_{L,R}^{s \tau}$ is real. 
Note that $x_L^{s \tau}/x_L^{b \tau} =\lambda$ is a typical reference value \cite{Cornella:2019hct}.
Here, favored parameter regions in the $\text{U}_1$ LQ model are exhibited on $x_L^{b\tau}$--$x_R^{b\tau}$ plane at the renormalization scale $\mu_{\text{LQ}} = m_{U_1}$.
The black contour represents the expected values of 
$R_{\Upsilon(3S)}$.
It is noted that if we adopt the $2\sigma$ constraint of $R_{\Upsilon(3S)}^{\rm exp}$ in Eq.~\eqref{eq:Upsilonexp},
the entire parameter region is allowed.
The blue and green regions can explain the $R_D$ and $R_{D^\ast}$ discrepancies 
within $1\sigma$, respectively.
The exclusion region by the LHC analysis ($\tau +\,$missing search) is outside the blue line, 
while the future prospect of the High Luminosity LHC (HL-LHC) is shown by the blue-dashed line,
see Table~\ref{tab:LHC_bound_one}.
Furthermore, a stronger collider bound comes from a non-resonant $\tau^+\tau^-$ search~\cite{ATLAS:2020zms,CMS:2022goy}, although it is model-parameter dependent. 
The outside of the red line is excluded by the non-resonant $\tau^+\tau^-$ search, where a public code \texttt{HighPT} \cite{Allwicher:2022mcg} is used.
The orange dashed line stands for a prediction in the case of the UV origin $\text{U}_1$ LQ with $\beta_R = -1 ~(\phi=\pi)$ as a reference value~\cite{Fuentes-Martin:2019mun}. 
The gray-shaded region is excluded by the $B_c$ lifetime, \ie, $\mathcal{B}(B_c \to \tau\overline\nu) > 0.6$.

From the figure, it is found that the current $R_{\Upsilon(3S)}^{\rm exp}$ overshoots favored parameter region from the $R_{D^{(*)}}$ anomalies.
The best-fit points of Eq.~\eqref{eq:LQfit_start} are shown by red crosses and predict $R_{\Upsilon(3S)}=0.9942$. 
Thus, the LFU violation in $R_{\Upsilon(nS)}$ is predicted to be very small in the $\text{U}_1$ LQ scenario.\\

%%%%%%%%%%%%%%%%%%%%%%
\begin{figure}[t]
    \centering
    \includegraphics[width=0.7\textwidth]{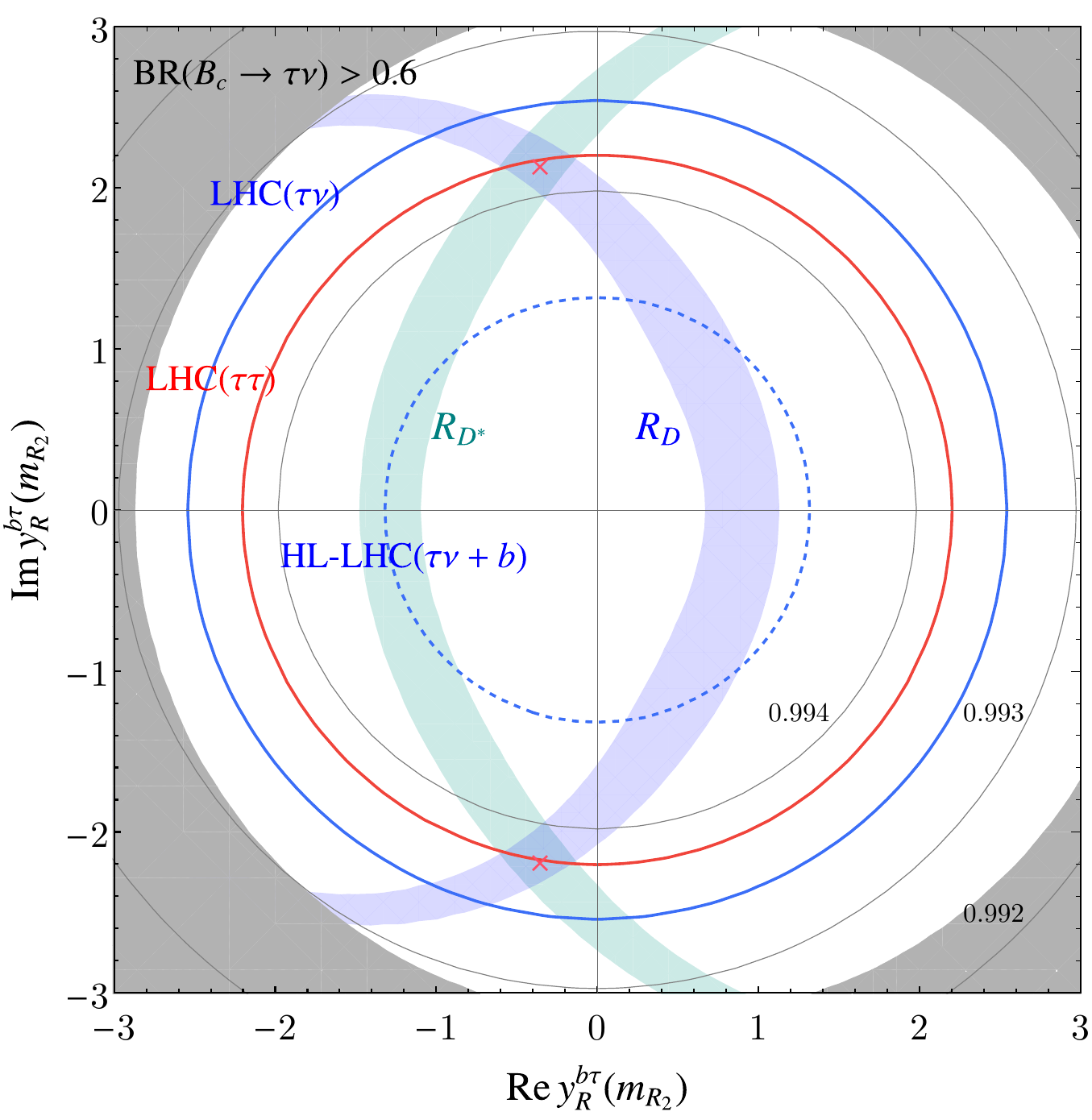}
    \caption{
    The correlation between $R_{\Upsilon(3S)}$ and $R_{D^{(\ast)}}$ in the $\text{R}_2$ LQ scenario with $C_{V_R}=0$ and   $m_{\text{R}_2}=2\,$TeV 
    by setting and $y_L^{c \tau}/|y_R^{b \tau} | =0.7 $.
    The color convention is the same as in Fig.~\ref{fig:U1Upsilon}.  
    The best-fit points of Eq.~\eqref{eq:LQfit_end} are shown by red crosses.
\label{fig:R2Upsilon}}
\end{figure}
%%%%%%%%%%%%%%%%%%%%%%

Next, we investigate the  $\text{R}_2$ LQ with $C_{V_R} =0$ scenario.
The $\text{R}_2$ LQ interaction with the SM fermions is given in Eq.~\eqref{eq:R2int}.
The generated charged current contributions are given in Eq.~\eqref{eq:R2charged}, while the neutral current one is
\beq
C_{VLR}^{b \tau} (\mu_{\rm LQ})= - \frac{|y_R^{b \tau}|^2}{2 m_{R_2}^2} \,.
\eeq
Since $C_{VLR}^{b \tau} < 0$, 
$R_{\Upsilon(3S)} / R_{\Upsilon(3S)}^{\rm SM}$ 
has to be less than $1$ again.

The result is shown in Fig.~\ref{fig:R2Upsilon}.
Here, we set 
 $m_{\text{R}_2}=2.0\,$TeV, 
 $y_L^{c \tau}/|y_R^{b \tau} | =0.7 $   with $|V_{cb}|=0.041$, 
 and take $y_R^{b \tau}(\mu_{\rm LQ})$ as complex value.
 Note that it is found in Ref.~\cite{Iguro:2023rom} that the choice of $y_L^{c \tau}/|y_R^{b \tau} | \approx 0.7$ can alleviate the bound from the non-resonant $\tau^+\tau^-$ searches. 
The color convention is the same as the $\text{U}_1$ LQ case.
The best-fit points in Eq.~\eqref{eq:LQfit_end} are shown by red crosses,  predicting $R_{\Upsilon(3S)}=0.9938$.
Similar to $\text{U}_1$ LQ interpretation, it seems crucial to measure $R_{\Upsilon(nS)}$ with $0.1\%$ accuracy in order to distinguish the $\text{R}_2$ LQ signal.

Note that the $\text{S}_1$ LQ does not contribute to $R_{\Upsilon(nS)}$, while the $\text{V}_2$ LQ can contribute. 
However we found that the $\text{V}_2$ contribution is also small; $0.1\%$ effect in $R_{\Upsilon(3S)}$ \cite{Iguro:2023prq}.

At the current stage, 
the large experimental uncertainty in $R_{\Upsilon(3S)}$ cannot allow a clear-cut conclusion.
One should note that the Belle and Belle II experiments have enough sensitivities to
the $R_{\Upsilon(nS)}$ measurements, which would be more accurate than the existing BaBar measurement~\cite{ishikawa}.

%%%%%%%%%%%%%%%%%%%%%%%%%%%%%
\section{Conclusions and discussion}
\label{sec:conclusion}
%%%%%%%%%%%%%%%%%%%%%%%%%%%%%

%%%%%%%%%%%%%%%%%%%%%%%%%%%%%%%
\begin{table}[t]
\centering
\newcommand{\bhline}[1]{\noalign{\hrule height #1}}
\renewcommand{\arraystretch}{1.5}
\rowcolors{2}{gray!15}{white}
%\addtolength{\tabcolsep}{5pt} % add space between columns
   \scalebox{0.95}{
  \begin{tabular}{lccc ccc cc} 
  %\toprule
  \bhline{1 pt}
  \rowcolor{white}
 &Spin & Charge & Operators&$R_D$ &$R_{D^*}$ 
 & LHC & Flavor 
 \\   %\begin{tabular}{c ccc} 
  \hline
  %%%%%%%%%%%%
 $H^\pm$ 
 &0&($\bf{1}$,\,$\bf{2}$,\,$\sfrac{1}{2}$)& $O_{S_L}$& \Large{\color{teal}{\bf{\checkmark}}} & \Large{\color{teal}{\bf{\checkmark}}} & $b\tau\nu$ &$B_c\to\tau\nu$, $F_L^{D^\ast}$, $P^{D^\ast}_\tau$, $M_W$\\
  %%%%%%%%%%%%
 $\text{S}_1$
 &0&($\bf{\bar{3}}$,\,$\bf{1}$,\,$\sfrac{1}{3}$)& $O_{V_L}$, $O_{S_L}$, $O_{T}$& \Large{\color{teal}{\bf{\checkmark}}} & \Large{\color{teal}{\bf{\checkmark}}}& $\tau\tau$ & $\Delta M_s$, $P_{\tau}^D$, $B \to K^{(\ast)} \nu \nu $\\
 %%%%%%%%%%%%
 $\text{R}_2^{(\sfrac{2}{3})}$ &0&($\bf{3}$,\,$\bf{2}$,\,$\sfrac{7}{6}$)&$O_{S_L}$, $O_{T}$, ($O_{V_R}$)& \Large{\color{teal}{\bf{\checkmark}}} & \Large{\color{teal}{\bf{\checkmark}}} & $b\tau\nu$, $\tau\tau$ &
 $P_{\tau}^{D^\ast}$, $M_W$, $Z \to \tau \tau$, $d_N$\\
 %%%%%%%%%%%%
 $\text{U}_1$ &1&($\bf{3}$,\,$\bf{1}$,\,$\sfrac{2}{3}$)&$O_{V_L}$, $O_{S_R}$& \Large{\color{teal}{\bf{\checkmark}}} & \Large{\color{teal}{\bf{\checkmark}}}& $b\tau\nu$, $\tau \tau$ & $\Delta M_s$, $R_{K^{(\ast)}}$, 
 $B_s \to \tau\tau$, $d_N$\\
 %%%%%%%%%%%%
 $\text{V}_2^{(\sfrac{1}{3})}$ &1&($\bf{\bar{3}}$,\,$\bf{2}$,\,$\sfrac{5}{6}$)&$O_{S_R}$& \Large{\color{teal}{\bf{\checkmark}}} & ${2\sigma}$& $\tau\tau$ & $B_s \to \tau\tau$, $B_u\to\tau\nu$, $M_W$\\
\bhline{1 pt}
%\bottomrule
   \end{tabular}
   }
%\addtolength{\tabcolsep}{-5pt} % set back to normal
    \caption{\label{tab:summary_table} 
Summary table for the single-mediator NP scenarios in light of the $b \to c \tau \nu$ anomaly.
We add implications for the LHC searches and flavor observables in the last two columns, which is useful to identify the NP scenario. 
In the $\text{V}_2^{(\sfrac{1}{3})}$ LQ scenario, $2\sigma$ for $R_{D^\ast}$ implies that it can explain the $R_{D^\ast}$ anomaly within the $2\sigma$ range 
(but not within $1\sigma$).
}
\end{table}
%%%%%%%%%%%%%%%%%%%%%%%%%%%%%%%

%%%%%%%%%%%%%%%%%%%%%%%%%%%%%
In this work, we revisited our previous phenomenological investigation and presented the statistical analysis of the LFU violation in $R_{D^{(*)}}$, 
including the new experimental data from the LHCb and Belle~II experiments. 
Starting with the re-evaluation of the generic formulae for $R_{D^{(*)}}$ by employing the recent development of the $B\to D^{(\ast)}$ transition form factors, 
we examined the NP possibility with the low-energy effective Lagrangian as well as the LQ models.
In addition to the constraints from the low-energy observables and the high-$p_T$ mono-$\tau$ search at LHC, 
the predictions on the relevant observables of $R_\Upsilon$, $R_{\Jpsi}$, and the $\tau$ polarizations $P_{\tau}^{D^{(*)}}$ are evaluated.

To be precise, we performed the $\chi^2$ fit to the experimental measurements of $R_{D^{(*)}}$ and the $D^*$ polarization $F_L^{D^*}$.
This updated analysis shows that the present data deviates from the SM predictions at $\sim 4\sigma$ level. 
Our fit result is summarized in Table~\ref{tab:fit:singlereal} with Eqs.~\eqref{eq:complex_start}--\eqref{eq:fit_CSL_complex_withBc} for the single-operator scenarios, 
and Table~\ref{tab:fit:leptoquark} with Eqs.~\eqref{eq:LQfit_start}--\eqref{eq:LQfit_end}, \eqref{eq:U2result} for the single-LQ scenarios. 
The NP fit improvements compared with the SM one are visualized by Pull as usual, and it is found that the SM-like vector operator still gives the best Pull. 

Due to the new LHCb and Belle~II results, the experimental world average has slightly come close to the SM predictions of $R_D$ and $R_{D^*}$.
Moreover, the recent $F_L^{D^*}$ result from the LHCb made the experimental value consistent with the SM prediction. 
These changes altered the previous situations that the scalar and tensor NP solutions to the $b \to c \tau \nu$ anomaly had been disfavored. 
Namely, the scalar and tensor NP interpretations have been revived now. 
On the other hand, it is found that the results of the LQ scenarios do not drastically change compared with the previous fit.

As it was pointed out in the literature, 
the precise measurements of the polarization observables $P_{\tau}^{D^{(*)}}$ and $F_L^{D^*}$ have the potential to distinguish the NP scenarios. 
In Figs.~\ref{fig:EFTprediction} and \ref{fig:LQprediction}, we show our predictions of $P_{\tau}^{D}$ and $P_{\tau}^{D^{*}}$ for the possible NP scenarios.
One can make sure that the single-operator NP scenario explaining the $b \to c \tau \nu$ anomaly can be identified by the $P_{\tau}^{D^{(*)}}$ measurements, 
which may be available at the Belle~II experiment.
On the other hand, the general LQ scenarios are hard to be distinguished due to predicting wide ranges of $P_{\tau}^{D^{(*)}}$. 
Once the LQ model with restricted interactions is constructed, however, 
we see that the $P_{\tau}^{D^{(*)}}$ measurement has significant potential to probe the LQ signature.
The high energy collider search is also important since the high-$p_T$ lepton search at the LHC can directly probe the NP interactions affecting the LFU ratios.

We also investigated the NP impacts on the LFU violation in the $\Upsilon(nS)$ decays. 
We found that the LFU ratio $R_\Upsilon$ is expected to be deviated from the SM prediction by $\mathcal{O}(0.1)\%$ in the $\text{U}_1$, $\text{R}_2$, and $\text{V}_2$ LQ scenarios for the $R_{D^{(\ast)}}$ anomaly, 
while $\text{S}_1$ LQ scenarios expect no deviation. 
Hence, an experimental accuracy of less than $0.1\%$ for the $R_{\Upsilon(nS)}$ measurement is necessary in order to identify the LQ scenario based on the distinct correlation.

In Table~\ref{tab:summary_table}, we put a summary check sheet to find which single-mediator NP scenarios are viable 
and to see important observables in order to identify the NP scenario responsible for the $b \to c \tau \nu$ anomaly. 
%

%%%%%%%%%%%%%%%%%%%%%%%%%%%%%%%%%%%%%%%%%%%%%%%%%%%%%%%%
\section*{Acknowledgements}
%%%%%%%%%%%%%%%%%%%%%%%%%%%%%
The authors would like to thank 
 Motoi Endo, 
 Akimasa Ishikawa,
 Satoshi Mishima,
 Yuta Takahashi,
 and
 Kei Yamamoto
 for fruitful comments and valuable discussion at different stages of the work.
%---------------------------------------------------------------------------
We also appreciate 
Monika Blanke,
Andreas Crivellin, 
Marco Fedele, 
Ulrich Nierste 
and 
Felix Wilsch for useful discussion.
%---------------------------------------------------------------------------
%%%
S.I.\ enjoyed the support from the Deutsche Forschungsgemeinschaft (DFG, German Research Foundation) under grant 396021762-TRR\,257.
S.I.\ thanks Karlsruhe House of Young Scientists (KHYS) for the financial support which enabled him to invite R.W.\ for the discussion.
%%%
The work of T.K.\ was supported by the Japan Society for the Promotion of Science (JSPS) Grant-in-Aid for Early-Career Scientists (Grant No.\,19K14706). 
%%%
The work of S.I.\ and T.K.\ are supported by the JSPS  Core-to-Core Program (Grant No.\,JPJSCCA20200002). 
%%%
R.W.\ was partially supported by the INFN grant `FLAVOR' and the PRIN 2017L5W2PT.
%%%
%---------------------------------------------------------------------------
%%%%%%%%%%%%%%%%%%%%%%%%%%%%%%%%%%%%%%%%%%%%%%%%%%%%%%%% 

\appendix
%%%%%%%%%%%%%%%%%%%%%%%%%%%%%%%
\section{Leptoquark interactions}
\label{sec:LQint}
%%%%%%%%%%%%%%%%%%%%%%%%%%%%%%%
The LQ interactions are classified with the generic $SU(3)_c \times SU(2)_L \times U(1)_Y$ invariant form \cite{Buchmuller:1986zs}. 
We leave details of the model constructions and then just introduce the interactions relevant for $b\to c\tau\overline\nu$.
As mentioned above, there are four viable candidates of LQs; 
$\text{U}_1,~\text{S}_1,~ \text{R}_2 $\cite{Angelescu:2018tyl} and $\rm{V}_2$~\cite{Iguro:2023prq}.
Their quantum numbers under $SU(3)_{\rm{C}}$, $SU(2)_{\rm{L}}$, $U(1)_{\rm{Y}}$ are summarized in Table \ref{tab:summary_table}.
%%%%%%%%%%%%%%%%%%%%%%%%%%%%%%%

First, the $\text{U}_1$ vector LQ interaction with the SM fermions, defined in the interaction basis, is given by 
\beq
\mathcal{L}_{U_1} =  x^{ij}_L \overline{Q}_i \gamma_{\mu} U^{\mu}_1 L_j 
+  x^{ij}_R \overline{d}_{Ri}\gamma_{\mu} U^{\mu}_1 \ell_{R j} 
+\textrm{h.c.}\,.
\label{eq:U1int}
\eeq
Integrating out the $\text{U}_1$ LQ mediator particle, then, the Wilson coefficients (WCs) for the charged current of our interest ($b \to c \tau \overline\nu$) is obtained as 
\beq
\begin{aligned}
C_{V_L} (\mu_{\rm LQ}) 
&=
\frac{1}{2 \sqrt{2} G_F V_{cb} } \frac{ (V x_L)^{c \tau} (x_L^{b \tau})^\ast}{m_{U_1}^2}\,,
\\
C_{S_R} (\mu_{\rm LQ}) &= 
-
\frac{1}{\sqrt{2} G_F V_{cb}}\frac{(V x_L)^{c \tau} (x_R^{b \tau})^\ast}{m_{U_1}^2}\,,
\label{eq:U1charged}
\end{aligned}
\eeq
where $V$ is the CKM matrix and the couplings $x_{L,R}$ are in the mass basis. 
The relative sign and factor two in Eq.~\eqref{eq:U1charged} come from the property of Fierz identity.

In a typical UV completed theory~\cite{Fuentes-Martin:2019mun}, 
the $\text{U}_1$ LQ is realized as a gauge boson generated from a large gauge symmetry and only couples to the third-generation SM fermions. 
Namely, $ x_R^{b \tau} =  x_L^{b \tau} \equiv g_U$, with the others to be zero, is indicated in the gauge interaction basis. 
Moving to the mass basis, then, generates a non-zero off-diagonal part such as $x_L^{c \tau}$ and also $x_R^{b \tau} = e^{-i \phi} x_L^{b \tau}$, 
where the phase comes from those in the rotation matrices to the mass bases of the left- and right-handed quark and lepton fields that are not canceled in general. 
Therefore, the UV completion of $\text{U}_1$ LQ suggests 
\begin{align}
    C_{S_R}  (\mu_{\rm LQ}) =  - 2 e^{i \phi} C_{V_L}  (\mu_{\rm LQ})\,, 
\end{align} 
as introduced in the main text. 
We also comment that an extension of the fermion families with a nontrivial texture of the fermion mass matrices is necessary to construct a practical UV model~\cite{Iguro:2021kdw}.

%%%%%%%%%%%%%%%%%%%%%%%%%%%%%%%

%%%%%%%%%%%%%%%%%%%%%%%%%%%%%%%
The $\text{S}_1$ scalar LQ interaction in the mass basis is given by  
\beq
\mathcal{L}_{{\rm S}_1} 
&=  \big{(}V^\ast y_L \big{)}^{ij}\, \overline{u^C_{L\,i}}\ell_{L\,j}{\rm S}_1-y_L^{ij}\,\overline{d^C_{L\,i}}\nu_{L\,j}{\rm S}_1+y_R^{ij}\, \overline{u^C_{R\,i}}\ell_{R\,j}  {\rm S}_1+ \mathrm{h.c.}\,.
\label{eq:S1L}
\eeq
In the scalar LQ scenario, the source of the generation violating couplings is off-diagonal element of  Yukawa matrices.  
Then the four-fermion interactions of $b \to c \tau {\overline \nu}$ are given by 
\beq
\begin{aligned}
C_{S_L} (\mu_{\rm LQ})
&= - 4 \, C_T (\mu_{\rm LQ})
= -\dfrac{1}{4 \sqrt{2} G_F V_{cb}}\dfrac{y_L^{b{\tau}} \big{(}y_R^{c\tau}\big{)}^\ast}{ m_{S_1}^2}\,, \\
C_{V_L} (\mu_{\rm LQ})
&= \dfrac{1}{4 \sqrt{2} G_F V_{cb}}\dfrac{y_L^{b \tau}\big{(}V y_L^\ast\big{)}^{c\tau}}{ m_{S_1}^2} \,.
\end{aligned}
\eeq
%%%%%%%%%%%%%%%%%%%%%%%%%%%%%%%

%%%%%%%%%%%%%%%%%%%%%%%%%%%%%%%
We also introduce the $\text{R}_2$ scalar LQ interaction.
$\text{R}_2$ is a SU(2) doublet and a component with $2/3$ of the electromagnetic charge ${\rm R}_2^{(\sfrac{2}{3})}$ can contribute to $b \to c \tau {\overline \nu}$.
The Yukawa interaction
\beq
\mathcal{L}_{{\rm R}_2}=  y_R^{ij} \, \overline{d}_{L\,i} \ell_{R\,j}\,{\rm R}_2^{(\sfrac{2}{3})}    +y_L^{ij} \overline{u}_{R\,i} \nu_{L\,j}\, {\rm R}_2^{(\sfrac{2}{3})} +\mathrm{h.c.}\,,
\label{eq:R2int}
\eeq
gives 
\beq
C_{S_L} (\mu_{\rm LQ})
&= 4 \, C_T(\mu_{\rm LQ}) 
= \dfrac{1}{4 \sqrt{2} G_F V_{cb}} \dfrac{y_{L}^{c \tau}\big{(}y_R^{b \tau}\big{)}^\ast}{m_{R_2}^2 }\,.
\label{eq:R2charged}
\eeq
In contrast to the above two LQ scenarios, the $\text{R}_2$ LQ does not generate $C_{V_L}$ but $C_{V_R}$. 
Thus we could expect solid predictions in polarization and related observables.
To generate $C_{V_R}$, indeed, a large mixing between two distinct $\text{R}_2$ LQ doublet is required to induce a proper electroweak symmetry breaking. 
See details in Refs.~\cite{Asadi:2019zja,Endo:2021lhi}.

Finally, we introduce the iso-doublet vector $\rm{V}_2$ LQ. 
A component with electromagnetic charge of $1/3$, $\rm{V}_2^{(\sfrac{1}{3})}$, contributes to $b\to c\tau\ov \nu$.
The interaction Lagrangian in the interaction basis is given by 
\begin{align}
{{\cal{L}}_{{\rm{V}}_2}}\,=\,-z_{L}^{ij}(\ov{d^{C}_{R\,i}}\gamma_\mu  \nu_{L\,j}){\rm{V}}_2^{(\sfrac{1}{3}),\mu}+(V^*z_{R})^{ij}(\ov{u^{C}_{L\,i}}\gamma_\mu  \ell_{R\,j})\rm{V}_2^{(\sfrac{1}{3}),\mu}+{\rm{h.c.}}\,,
\label{Eq:fermoin_VQ2}
\end{align}
where indices $i,\,j$ are labels of flavor.
Integrating out the $\rm{V}_2$ gives 
\begin{align}
C_{S_R}=-\frac{1}{\sqrt{2}G_F V_{cb}}\frac{z_L^{b\tau}(Vz_R^*)^{c\tau}}{m_{\rm{V}_2}^2}\,.
\label{eq:CSR_general}
\end{align}
It is noted that although the $\rm{U}_1$ LQ scenario also contributes to $C_{S_R}$, 
the scenario can have $C_{V_L}$ solely so that the $R_{D^{(*)}}$ anomaly can be explained.
Furthermore,
the isospin partner $\rm{V}_2^{(\sfrac{4}{3})}$
provides distinct model predictions compared to the $\rm{U}_1$ LQ scenario~\cite{Iguro:2023prq}.

%%%%%%%%%%%%%%%%%%%%%%%%%%%%%%%%%%%%%
%%%%%%%%%%%%%%%%%%%%%%%%%%%%%%%%%%%%%

\bibliographystyle{utphys28mod}

\bibliography{ref}

\end{document}